\documentclass[journal]{IEEEtran}

\ifCLASSINFOpdf
\else
\fi

\usepackage{cite}
\usepackage{url}
\usepackage{multicol}
\usepackage{multirow}
\usepackage{array}
\usepackage{booktabs}
\usepackage{amsmath}
\usepackage{algorithmic}
\usepackage{tablefootnote}
\usepackage{threeparttable} 
\usepackage{subfigure} 
\usepackage{graphicx}
\usepackage{makecell}
\usepackage{color}
\usepackage{url}
\begin{document}

\title{Joint Communication and Sensing toward 6G: Models and Potential of Using MIMO}

	\author{Xinran~Fang,~\IEEEmembership{Student Member,~IEEE,}
	Wei~Feng,~\IEEEmembership{Senior Member,~IEEE,}
	Yunfei~Chen,~\IEEEmembership{Senior Member,~IEEE,}\\
	Ning~Ge,~\IEEEmembership{Member,~IEEE,}
	and Yan~Zhang,~\IEEEmembership{Fellow,~IEEE}
	\thanks{Xinran~Fang, Wei~Feng (Corresponding author), and Ning~Ge are with the Beijing National Research Center for Information Science and Technology, Department of Electronic Engineering, Tsinghua University, Beijing 100084, China (e-mail: {fxr20}@mails.tsinghua.edu.cn, {fengwei}@tsinghua.edu.cn,  {gening}@tsinghua.edu.cn).
	
	Yunfei~Chen is with the School of Engineering, University of Warwick, Coventry CV4 7AL, U.K. (e-mail: {Yunfei.Chen}@warwick.ac.uk). 
	
	Yan Zhang is with Department of Informatics,
	University of Oslo, Norway. (e-mail: {yanzhang}@IEEE.org).}
}
\maketitle

\begin{abstract}
The sixth-generation (6G) network is envisioned to integrate communication and sensing  functions, so as to improve the spectrum efficiency (SE) and support explosive novel applications. Although the similarities of wireless communication and radio sensing lay the foundation for their combination, there is still considerable incompatible interest between them.  To simultaneously guarantee the communication capacity and the sensing accuracy, the multiple-input and multiple-output (MIMO) technique plays an important role due to its unique capability of spatial beamforming and waveform shaping. However, the configuration of MIMO also brings high hardware cost, high power consumption, and high signal processing complexity. How to efficiently apply MIMO to achieve balanced communication and sensing performance is still open. In this survey, we discuss joint communication and sensing (JCAS) in the context of MIMO. We first outline the roles of MIMO in the process of wireless communication and radar sensing.  Then, we present current advances in both communication and sensing coexistence and integration in detail. Three novel JCAS MIMO models are subsequently discussed by combining cutting-edge technologies, i.e., cloud random access networks (C-RANs), unmanned aerial vehicles (UAVs) and reconfigurable intelligent surfaces (RISs). Examined from the practical perspective, the potential and challenges of MIMO in JCAS are summarized, and promising solutions are provided. Motivated by the great potential of the Internet of Things (IoT), we also specify JCAS in IoT scenarios and discuss the uniqueness of applying JCAS to IoT.  In the end, open issues are outlined to envisage a ubiquitous, intelligent and secure JCAS network in the near future.

\end{abstract}

\begin{IEEEkeywords}
communication and sensing coexistence, communication and sensing integration, joint communication and sensing (JCAS), multiple-input and multiple-output (MIMO), radar sensing.
\end{IEEEkeywords}

%
\IEEEpeerreviewmaketitle

\section{Introduction}

\IEEEpubidadjcol
Combining communication and sensing in wireless networks has recently attracted great attention. It not only allows for more efficient spectrum usage but also provides efficient dual-function services for many applications, e.g., intelligent transportation \cite{89}, smart factories \cite{88}, and the Internet of Things (IoT) \cite{87}. This has made joint communication and sensing (JCAS) a promising candidate for future networks.

The early motivation of JCAS comes from the scarcity of spectrum resources \cite{90}. With the increasing requirements of high-resolution sensing and high-rate communication, communication and sensing systems are constantly expanding and merging their frequency bands. For example, it has been reported by \cite{82} that the global system for mobile communication shares the same spectrum with high UHF radars and that the long-term evolution (LTE) and the WiMax system partially occupy the spectrum of S-band radars. In addition, for the shake of the efficient usage of the wide bandwidth,  spectrum sharing is also extended to the mmWave band \cite{91}. Serious mutual interference motivates  communication and sensing systems to cooperate.

Since wireless communication and radio sensing both use radio signals to carry information, the idea of integrating them into one platform naturally arises. Such an integrated  communication and sensing system  has incomparable benefits of low cost, low power consumption and compact volume. This brings new opportunities to those applications that require both communication and sensing services, but their platforms are incapable of supporting the both. To achieve this kind of JCAS, researchers have made considerable efforts. The radar-centric schemes try to use  radar platforms to achieve communication functions. The communication-centric schemes try to extract target information from communication signals. To achieve balanced communication and sensing performance, devising a novel dual-function waveform has been proposed and investigated. 

Despite these fruitful results, there is still a long way to go for practical JCAS deployments. Inherent difference between communication and sensing makes the joint design difficult.  Looking back to the development of  wireless communication and radio sensing, the multiple-input and multiple-output (MIMO) technique
plays an important role in their progress. MIMO extends resource utilization into the spatial domain, and thus greatly improves the communication rate and sensing resolution. There is no doubt that MIMO would also be a strong supporter for JCAS. However,  MIMO inherently brings the dimensional problem in terms of the overhead and processing complexity. How to efficiently use MIMO to build a satisfying JCAS network has not yet been fully discussed.
\begin{table*}[htbp]
	
	\centering
	\caption{State of art surveys on JCAS}
		\label{table0-0}
		\begin{threeparttable}
			\begin{tabular}{|p{2cm}|p{0.7cm}|p{2cm}|p{3cm}|p{8cm}|}
				\hline
				\textbf{Paper}&\textbf{Year}&\textbf{Topic}\tnote{1}&\textbf{Focus}&\textbf{Main contribution}\\ 
				\hline
				Han \emph{et al.} \cite{95}&2013&C\&S integration&system prototype and performance&A survey specialized in the dual-function system, the system architecture and performance under different waveforms are the main covered issue\\
				\hline
				Hassanien \emph{et al.} \cite{93,94}&\makecell[l]{2016,\\2019}&C\&S integration &embedding schemes&A survey specialized in signaling strategies of radar-centric C\&S integration\\
				\hline
				Zheng \emph{et al.} \cite{90}&2019&C\&S coexistence and integration&communication and sensing coexistence &A  survey specialized in C\&S coexistence, covering three typical spectrum sharing scenarios\\
				\hline
				Mishra \emph{et al.} \cite{91}&2019&C\&S coexistence and integration&signal processing&A  survey specialized in mmWave JCAS, mainly reviewing the mmWave characteristics and signal processing techniques for C\&S coexistence and co-design\\
				\hline
				Feng \emph{et al.} \cite{r12}&2020&C\&S coexistence and integration&comprehensive overview&A comprehensive survey on the state-of-the-art JCAS, from coexistence, cooperation, co-design to collaboration\\
				\hline
				Liu \emph{et al.} \cite{82}&2020&C\&S coexistence and integration&detailed working regime &An overview of the state-of-the-art techniques and a detailed case study on the working regime of the DFRC\\
				\hline
				

				Zhang \emph{et al.} \cite{84}&2021&C\&S integration&signal processing&A survey specialized in the signal processing of C\&S integration, with balanced coverage of T\&R techniques \\
				\hline
				
				Zhang \emph{et al.} \cite{83}&2021&C\&S integration&perceptive mobile network&A comprehensive survey on the perceptive mobile network, mainly covering the issues of framework design, system evolution and key technologies\\
				\hline

				Wild \emph{et al.}  \cite{86}&2021&C\&S integration&cellular-based C\&S integration   &A survey on specialized in cellular-based C\&S integration, with the focus on the  issues of waveform candidates,  parameter selections and resource allocation \\
				
				\hline
				Cui \emph{et al.} \cite{r13}&2021&C\&S integration&IoT scenarios&A macroscopic description of the motivations, applications, trends and challenges of JCAS in IoT \\
				
				\hline
				Liu \emph{et al.} \cite{85}&2022&C\&S integration&comprehensive overview&A comprehensive survey on C\&S integration, mainly including the issues of background, key applications and state-of-the-art
				approaches\\
				
				\hline
				Liu \emph{et al.} \cite{r14}&2022&C\&S integration&fundamental limits&A specialized survey on the fundamental limits of sensing and C\&S integration, including the device-free and device-based cases\\
				\hline 
				This survey&2022&C\&S coexistence and integration& JCAS MIMO&A survey specialized in JCAS of using MIMO, discussing basic models, potential and challenges of JCAS MIMO designs \\
				\hline
			\end{tabular}
		\begin{tablenotes}
			\footnotesize
			\item[1] C\&S: communication and sensing.
		\end{tablenotes}

		\end{threeparttable}
\end{table*}
\IEEEpubidadjcol
In the past decade, many outstanding studies have reviewed JCAS from different aspects. To compare this survey with other related works  \cite{90}--\cite{r14}, we summarize their topics, main focuses and contributions in Table \ref{table0-0}. As we can see, although these surveys have provided us with a clear map of JCAS development, none of them paid particular attention to the role of MIMO in JCAS. As a key enabler of JCAS, the potential and challenges of MIMO are complementary. 
It is necessary to clarify the basic models and schemes of MIMO such that it can be efficiently used for JCAS. This would further help us figure out the fundamental trade-off between communication and sensing.
Therefore, in this paper, we discuss JCAS specialized in the MIMO context. We present recent advances in both communication and sensing coexistence and communication and sensing integration. In addition, we investigate the usage of MIMO combined with cutting-edge techniques, i.e., cloud radio access networks (C-RANs), unmanned aerial vehicles (UAVs) and the reconfigurable intelligent surfaces (RISs). With the principle of simplicity and robustness, we summarize the potential and challenges that MIMO brings and outline promising solutions in each subsection. We also pay our attention to the JCAS deployment in IoT scenarios. By considering the traits of IoT devices, i.e., ubiquitously distributed, cheaply produced and power-limited, we discuss the issues of ubiquity, green, complexity and cooperation for JCAS in IoT. Based on the lessons learned, open issues are presented in the end to envisage a ubiquitous, intelligent and secure JCAS network in the near future.

In this paper, the term ``MIMO" refers to using multiple RF chains at transmitters,  receivers or both sides. We also consider the exception of phased array radar, which only has one RF chain and the RF chain is connected with many radiating elements. The remainder of this paper is organized as follows. In Section \uppercase\expandafter{\romannumeral2},  a general review of the communication and sensing evolution accompanied by  MIMO advancements is presented, and the roles that MIMO plays are outlined. In Section \uppercase\expandafter{\romannumeral3}, we provide a detailed introduction of current advancements in JCAS with an emphasis on the potential and challenges that MIMO brings. Section \uppercase\expandafter{\romannumeral4} discusses three novel JCAS MIMO models, i.e., cooperative MIMO, dynamic three-dimensional (3D) MIMO, and joint active and passive MIMO. Section \uppercase\expandafter{\romannumeral5} specifies JCAS in IoT scenarios, Section \uppercase\expandafter{\romannumeral6} outlines open issues, and Section \uppercase\expandafter{\romannumeral7} draws the conclusions.

\section{Review of the Potential of MIMO Techniques}

\begin{figure*}[!htbp]
	\centering	
	\includegraphics[width=0.7\textwidth]{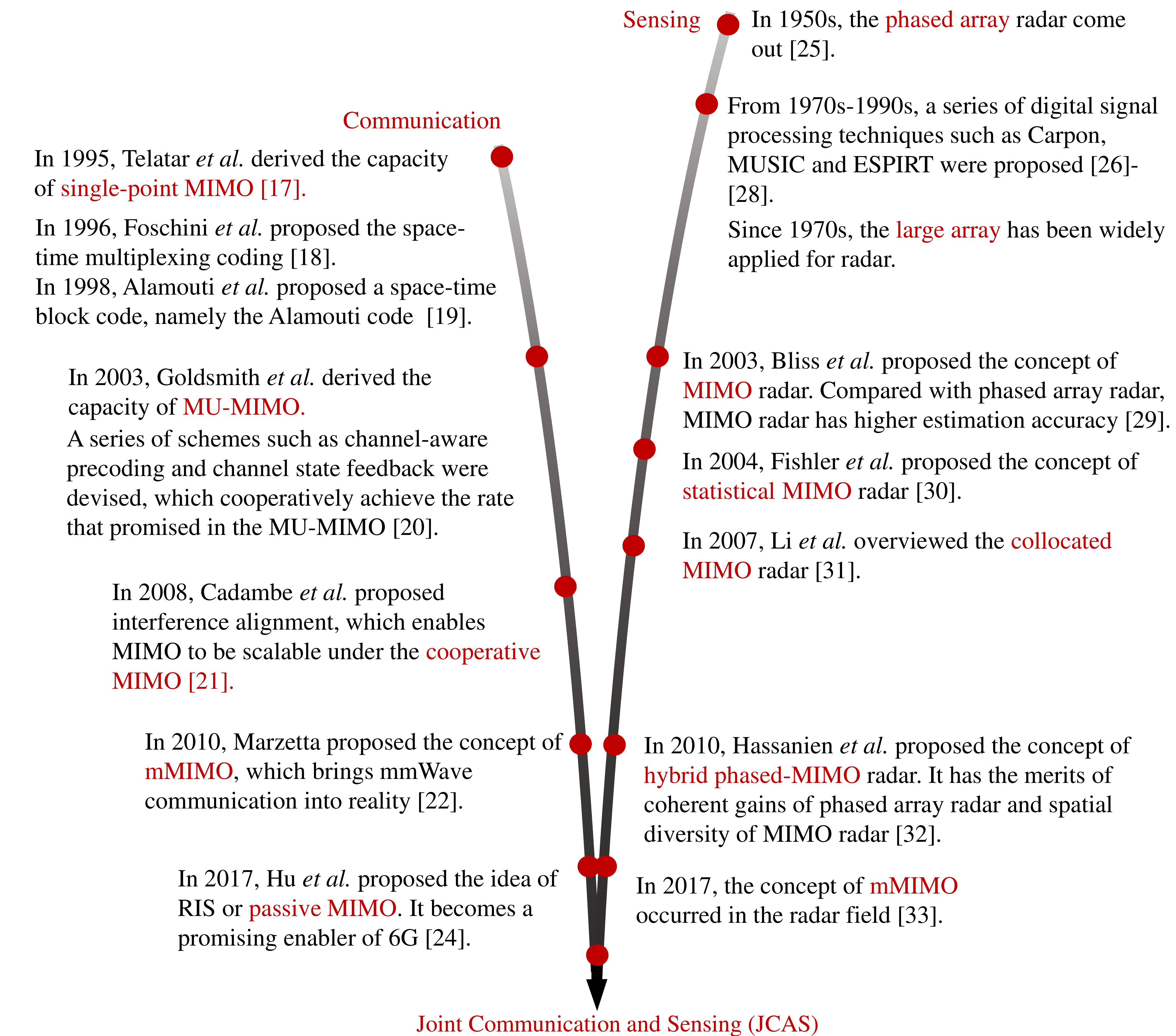}
	\caption{Key milestones in the communication and sensing evolution accompanied by MIMO advancements.}
	\label{fig}
\end{figure*} 

\subsection{Key MIMO Advancements in Communication and Sensing Evolution}

As shown in Fig. \ref{fig}, we summarize the key milestones of communication and sensing since the birth of MIMO. In the 1990s, the capacity of single-point MIMO was calculated \cite{144}. The striking capacity improvement over  single-antenna systems thus kicked off the era of MIMO communication. Subsequently, two outstanding space-time codes, termed V-BLAST/D-BLAST and Alamouti code, achieved the full utilization of spatial multiplexing and spatial diversity \cite{146}\cite{147}. These breakthroughs shifted MIMO from the theoretical concept into reality. Subsequently, the single-user MIMO  (SU-MIMO) was extended to the multi-user MIMO (MU-MIMO).  In addition
to using spatial degrees of freedom (DoFs) to obtain the processing gains, MU-MIMO supports the spatial sharing among different channels. A series of practical strategies, including channel-aware precoding and multi-user reception, were devised. Combined with effective scheduling schemes, the achieve rate of MU-MIMO approaches the theoretical limit \cite{140}. 
Moving forward, people extended MU-MIMO to the cooperative MIMO, where distributed transceivers cooperate with each other to exploit the spatial efficiency. 
A creative idea named interference alignment was proposed, which achieves the rate of $K/2\log(1+\text{SNR})$  when serving $K$ users \cite{142}. The gap between the sum capacity of SU-MIMO and the rate of cooperative MIMO was greatly narrowed. 
In the last fifteen years, massive MIMO (mMIMO) has occurred. Compared with MIMO, mMIMO uses a much larger array. The quantitative change in antenna numbers brings amazing qualitative changes. Research shows that if the transmitter uses an infinite number of antennas to serve only a few users, the randomness of fast fading would vanish, and the channels tend to be orthogonal \cite{149}. As a result, the inter-user interference disappears,  and the scalability of MIMO is greatly improved. Moreover, the coherent processing of mMIMO could combat the high path loss of mmWave channels. By leveraging the sparsity of mmWave channels, the techniques of hybrid digital-analog precoding and compressed sensing were ingeniously applied to overcome the high cost of mMIMO \cite{155}. The combination of mMIMO and mmWave communication promoted the progress of wireless communication.  
Just when the mMIMO is still under development, a brand-new technique, named RIS, or passive MIMO, has come into the research vision \cite{157}. 
Specifically, the RIS is a surface that consists of many low-cost and energy-efficient adjustable units. By adjusting these units, the RIS works to reform the incident signal into a desired output.
Since the unit of RISs is nearly passive and low-cost, there are usually massive units in one RIS. This brings great DoFs and enables RISs to enhance the communication link in an energy-efficient way. Due to the outstanding merits of low cost and high DoFs, the RIS is a popular candidate for future 6G.

For radars, the usage of the antenna array dated back to the 1950s \cite{n126}. At that time, people invented the first phased array radar, which overcomes the drawback of low scanning frequency and limited scanning range of mechanical scanning radars. 
Subsequently, with the progress of digitization,  a plethora of processing algorithms, e.g., Capon, MUSIC, ESPIRT, etc., have been proposed \cite{128,129,130}. Large phased arrays that consist of hundreds of radiating elements have become increasingly common. They brought improved spatial resolution for sensing.
Moving forward, the MIMO concept was introduced from the communication field to the sensing field. Thanks to the great DoFs, MIMO brings better sensing performance compared with the phased array \cite{150}. Subsequently, the statistical MIMO radar was developed, which uses widely spaced transmit and receive (T\&R) antennas to observe targets from different perspectives \cite{151}. In addition,  MIMO radar with collocated antennas, which allows the correlation among different antennas, was also proven to be excellent \cite{152}. By exploiting waveform diversity, this kind of radar is characterized by super resolution and super parameter identifiability. To exploit both the coherent gains of phased array radars and the waveform diversity of MIMO radars, the concept of hybrid phased-MIMO radars was proposed, which coincides with the idea of hybrid digital-analog precoding in the communication field \cite{153}. More recently, researchers have also extended MIMO into mMIMO to further pursue high-quality sensing \cite{154}. 

\subsection{Merits of MIMO in Communication Systems}

\paragraph{Spatial diversity}  Spatial diversity is mainly used to combat fading, which averages the deep fading probability from one path to multiple independent paths. 
Related studies have proven that if the  antenna spacing is more than ten wavelengths, signals transmitted or received by different antennas experience independent fading. MIMO provides a maximum of $M_c*N_c$ \footnote{$M_c$ and $N_c$ are the numbers of transmit and receive antennas of the communication transceiver.} independent T\&R  paths.  Using this property, if transmit antennas emit the same signal, the receiver thus obtains $M_c*N_c$ independent replicas. The probability that these replicas all go through deep fading is the $M_c*N_c$ multiplications of the probability of one path. Therefore, the outage probability is largely reduced \cite{96}.   

\paragraph{Spatial multiplexing}
Spatial multiplexing is mainly used to  increase spectrum efficiency (SE). In contrast to spatial diversity, which aims to combat fading, spatial multiplexing utilizes fading to improve the system rate \cite{96}. If  T\&R pairs experience independent fading, the corresponding channel matrix is more likely to be well conditioned and  full rank. This channel can thus be decomposed into multiple parallel paths. Using this property, multiple streams  are allowed to be transmitted concurrently using overlapped time and frequency resources. 

\paragraph{Flexible beamforming}
With multiple choices to exploit spatial diversity and spatial multiplexing, MIMO supports different beamforming strategies for different scenarios \cite{97}. Take  transmit beamforming as an example.  When serving one user, the high directional beam is used to improve the received signal-to-noise ratio (SNR).
When serving multiple users, spatially separable beams are applied  to support multi-stream transmissions.
Concerning co-channel interference,  zero-forcing precoding provides interference-free channels for different users.
When in a cellular system, the pencil beam limits the power leakage to adjacent cells, and thus, the background noise is controlled. In essence, different beamforming schemes exploit different degrees of usage of spatial diversity and spatial multiplexing.
\subsection{Merits of MIMO in Sensing Systems}
\paragraph{Spatial diversity}
Similar to communication, one of the outstanding merits of MIMO radar is spatial diversity \cite{98}. Using properly spaced antenna arrays, $M_r*N_r$ \footnote{$M_r$ and $N_r$ are the numbers of transmit and receive antennas of the radar.} independent observations could be obtained at the receiver. This reduces the probability of missing targets and makes the detection robust to cluttering effects.
From another perspective, the collocated MIMO radar, which allows correlations among different T\&R paths, has great parameter identifiability and high resolution \cite{99}. Compared with the phased array radar, MIMO radars could improve the parameter identifiability at most $M_r$-fold.  

\paragraph{Adaptive waveform manipulation} 
Another unique feature of the MIMO radar is the capability of adaptive waveform designs \cite{99}. To improve the reliability of target detection, the probing beam could emit high power toward the target while nulling surrounding clutters. To improve the multi-target identifiability,  correlations among different sub-beams could be surpassed by adjusting the transmit covariance matrix.  To detect the high-speed target, which is likely to be missed by single-beam scanning, multiple directional beams could be simultaneously generated to illuminate the whole area. In regard to target tracking, the dynamic beam is used to follow this target. Similar to communication, different waveform designs are achieved by the different degrees of usage of
spatial diversity and spatial multiplexing.  

\subsection{Summary} In Table \ref{tab2}, we summarize the main role of MIMO and the conflicts of using MIMO in these two functions.
As we can see, the exploitation of spatial resources and supporting flexible adaptations are the main merits of MIMO. 
Its roles in communication and sensing are similar in many aspects, e.g., using the spatial diversity to improve the reliability. Despite this,  communication and sensing have their own emphasis on the role of  MIMO. 
Communication mainly uses MIMO to adapt to channels, with the emphasis on directional beamforming, so as to point to different users and control interference. Radars mainly use MIMO to adapt to targets, with the emphasis on the waveform, so as to facilitate signal processing and target information extraction. Due to different emphases, the JCAS MIMO designs need to balance their fundamental interests. In the next section, we present current advancements in JCAS MIMO and detail this issue.
\begin{table*}[t]
	\centering
		\caption{MIMO enabled communication and sensing system}
			\label{tab2}
		\begin{tabular}{|p{2cm}|p{2.5cm}|p{3.2cm}|p{4cm}|p{4cm}|}
			\hline
			\textbf{System}&\textbf{Metric}&\textbf{The Role of MIMO}&\textbf{Function}&\textbf{Incompatibility}\\
			\hline
			\multirow{5}{*}{communication}&single-user rate&coherent beamforming&improve the directionality and SNR&\multirow{5}{3.9cm}{
				may cause high PAPR, high sidelobe and destroy the correlation properties}\\
			\cline{2-4}
			&outage probability&space-time coding&combat multi-path fading, improve the reliability&\\
			\cline{2-4}
			&system capacity&spatial precoding&construct spatial independent channels, lower down the interference&\\
			\hline
			\multirow{5}{*}{sensing}&detection probability&coherent beamforming&improve the SNR&\multirow{6}{3.9cm}{
			influence the spatial separability, may cause symbol distortion and influence the data carrying capability}\\
			\cline{2-4}
		    &detection reliability&waveform diversity&lower down the target missing probability&\\
			\cline{2-4}
			&estimation resolution&waveform shaping&lower down the sidelobe, ensure correlation properties, constant envelop, etc.&\\
			\hline
		\end{tabular}
\end{table*}

\section{State-of-the-art of MIMO-Empowered JCAS}

In the literature, the convergence of communication and sensing is a gradual process. At first,  spectrum scarcity motivates the two to coexist. 
Further advancements in fabrication and signal processing open the door to communication and sensing integration.
In the following, we first review the coexistence of communication and sensing and then review the integration of communication and sensing. We put our focus on the communication and sensing trade-off using MIMO. The main roles that MIMO plays in the dual-function design are summarized, and promising solutions are outlined.

\subsection{Communication and Sensing Coexistence}   
\begin{table*}[t]
	
	\centering
	\caption{MIMO enabled communication and sensing coexistence}
	\label{table2}
	\begin{threeparttable}
	\begin{tabular}{|p{2.8cm}|p{2cm}|p{3.3cm}|p{2.3cm}|p{1cm}|p{1.2cm}|p{2.5cm}|}
		\hline
		
		\textbf{Setting}\tnote{1}&\textbf{Interference}\tnote{2}&\textbf{Optimization Variable}\tnote{2}&\textbf{Channel Model}&\textbf{Perfect CSI}&\textbf{Same Sampling Rate}&\textbf{Programming Algorithm}\\ 
		\hline
		ST-MIMO, MISO&C to S&BS precoding&unspecific&yes&no&LCVM\cite{1}\\
		\hline
		ST-MIMO, MU-MISO&C\&S, inter-user&BS precoding&flat
		Rayleigh fading&yes/no&yes&SDP\cite{2}\\
		\hline
		ST-MIMO, MU-MISO&C\&S, inter-user&BS precoding&flat Rayleigh fading& yes/no&yes&gradient projection \cite{3}\\
		\hline
		ST-MIMO, FD MU-MISO&C\&S, inter-user, UL\&DL&joint BS DL precoding and UL power allocation&flat Rayleigh fading&yes/no&yes&SOCP\cite{4}\\
		\hline
		ST-MIMO, MU-MISO&C\&S, inter-user&joint BS and RIS precoding&block fading and
		quasi-static&yes&unspecific&SDP\cite{5}\\
		\hline
		ST-MIMO, MU-MIMO&S to C&radar precoding&flat Rayleigh  fading&yes&no&singular value decomposition (SVD)\cite{7}\\
		\hline
		ST-MIMO, multiple MU-MIMO&S to C&radar precoding&block fading and quasi-static&yes&yes&SVD	
		\cite{47}\\
		\hline
		MT-MIMO, MU-MIMO&S to C&radar precoding&block fading& yes&yes&SDP\cite{6}\\
		\hline
		MIMO, unspecific&S to C&radar waveform&unspecific&no&no& quadratic programming \cite{8}\\
		\hline
		ST-MIMO, SU-MIMO&C\&S&joint radar and BS precoding&block fading&yes&yes&SOCP\cite{10}\\
		\hline
		ST-MIMO, SU-MIMO&C\&S&joint radar and BS precoding&block fading&yes&no&alternating optimization and sequential convex programming\cite{67}\\
		\hline
		ST-MIMO, SU-MIMO&C\&S&S: waveform, receive filter, C: space-time code matrix&flat fading  &yes&yes&SOCP \cite{18}\\
		\hline
		MT-MIMO, SU-MIMO&C\&S&joint radar waveform and communication code book&unspecific&yes&yes&SDP\cite{112}\\
		\hline
		ST-MIMO, FD MU-MIMO&C\&S, inter-user, UL\&DL&S: precoding, C:UL\&DL T\&R precoding&composite channel model&no&no&SVD,SDP\cite{11}\\
		\hline
		MT-MIMO, SU-MIMO&C\&S&S: subsampling matrix, C: transmit covariance matrix&flat Rayleigh fading&yes&yes/no& Lagrangian dual decomposition\cite{12}\\
		\hline
		MT-MIMO, SU-MIMO&C\&S&S:transmit covariance matrix, subsampling matrix,	C: transmit covariance matrix&flat and block fading&yes&yes&SDP\cite{13}\\ 
		\hline
		MT-MIMO, MU-MIMO&C\&S, inter-user&joint radar and BS, T\&R precoding&unspecific&yes&yes&SDP \cite{17}\\
		\hline
		ST-MIMO, multiple SU-MISO&C\&S, inter-cell&joint radar waveform and BS precoding&flat fading&no&yes&ADMM,SDP\cite{108}\\
		\hline 
		MT-MIMO, MU-MIMO&C\&S, inter-user&joint radar and BS, T\&R  precoding&block fading and quasi-static&yes&yes&alternating optimization \cite{14}\\
		\hline
		ST-MIMO, MU-MIMO&C\&S, inter-user&joint radar and BS transmit precoding&time varying and frequency selective&yes&yes&linear equations
		\cite{15}\\
		\hline
		ST-MIMO, SU-MIMO&C\&S&joint radar transmit power, receive filter and communication transmit covariance&unspecific&yes&yes&block coordinate ascent method \cite{75}\\
		\hline
	\end{tabular}
\begin{tablenotes}
	\footnotesize
	\item[1] C\&S: communication and sensing, ST-MIMO: single-target MIMO, MT-MIMO: multi-target MIMO, SU-MISO: signle-user multi-input and single-output (MISO), MU-MISO: multi-user MISO.
	\item[2] C to S: communication to sensing, C: communication, S: sensing.
	\end{tablenotes}
\end{threeparttable}
	\label{TABLE1}
\end{table*}
Spectrum sharing is a key topic addressed by communication and sensing when their respective systems are separate.
The mutual interference  motivates originally separate systems to cooperate. The term ``coexistence" is used to depict the relationship of communication and sensing  in this situation. Spectrum sharing between communication and sensing is different from that in traditional communication systems. First, the radio sensing system referring to radars is usually characterized by high-power emissions \cite{82}.  If the communication transceiver is directly interfered by radar signals, corresponding data transmissions inevitably experience a fatal outage. As for radar sensing, besides SNR, particular requirements on the waveform, e.g., the constant envelope and good correlation properties \cite{17}, are equally important. 
To achieve peaceful coexistence,  there is at least one participant to make adaptations.   In the following, current studies are divided into communication adaptation schemes, radar adaptation schemes, and joint adaptation
schemes and are reviewed. 
\subsubsection{Communication Adaptation Schemes}
Communication adaptation schemes refer to the schemes where only communication systems make adaptations while radar systems remain unchanged.
Ebtihal H. G. \emph{et al.} considered a  one-user and one-target scenario and applied linear constraint variance minimization (LCVM) beamforming  to the base station (BS) so that the user is allowed to reuse the radar spectrum \cite{1}. In regard to multi-user scenarios,  both inter-user and inter-system interference were considered \cite{2,3,4}. Liu \emph{et al.} first proposed a robust beamforming scheme to maximize the radar detection performance under the constraint of the user rate \cite{2}. Then,  the authors found it is not power-efficient to totally eliminate the inter-user interference. The interference could be constructive if it helps align the received symbols in the right discrimination region. A precoding scheme was proposed to utilize the inter-user interference as the constructive green power rather than harmful noises \cite{3}. Keshav \emph{et al.} designed DL beamforming and uplink (UL) power allocation for a full duplex (FD) communication system \cite{4}. In this study, the two-tier spectrum sharing of uplink and downlink (UL\&DL) and communication and sensing achieves high SE. But more computing resources were required for solving the high-dimensional problem. More recently, Wang \emph{et al.} proved the effectiveness of RISs for interference mitigation using a joint BS transmit precoding and RIS passive precoding scheme. In this design, the RIS impact on radars was modeled the same as clutters. But the adjustment of the RIS would change radar echoes. This dynamic was not evaluated in this study \cite{5}.


\subsubsection{Radar Adaptation Schemes}
Schemes that adjust radar systems pay great attention to the radar waveform,  because a good waveform is the key to high-quality sensing. To control the interference to the communication system, Alireza \emph{et al.} designed a zero-forcing precoding that projects radar interference into the null space of the effective interfering channels. However, this leads the radar waveform far away from the desired patterns. The authors further relaxed the condition of null-space projection so that more DoFs were reserved for waveform shaping \cite{7}. Mahal \emph{et al.} further applied this scheme for the coexistence of a radar and a coordinated multi-point cellular system \cite{47}. Dan \emph{et al.} divided radar antennas into several subarrays to exploit the high directivity of phased array radars and the waveform diversity of MIMO radars. The radar SNR was maximized under the user-rate constraint \cite{6}. Hai \emph{et al.} devised a reception scheme that combines target focusing and interference mitigation into the beamforming design \cite{48}. In addition to spatial beamforming, Kang \emph{et al.} proposed a precoding scheme that  considers not only the spatial shape but also the spectrum shape of the radar waveform. A spectrum constraint was devised to control the radar power in the band that communication users occupy \cite{8}.  Moreover, Kilani \emph{et al.} investigated a placement scheme of distributed radars. Different from previous studies that directly regard communication signals as interference, the authors applied the successive interference cancellation (SIC) technique, which
first decodes the message of communication users and then removes it from the received signals. To ensure radars decode communication messages successfully, the placement design considered the proper position to both detect targets and decode messages \cite{9}.    

\subsubsection{Joint Adaptation Schemes}
In terms of joint optimization, Li \emph{et al.} considered a clutter environment and optimized the communication and radar transmit precoding to maximize the signal-to-interference-plus-noise ratio (SINR) of the radar. To avoid the high complexity of semidefinite programming (SDP), an efficient second-order cone programming (SOCP) algorithm was proposed for the radar precoding \cite{10}. In \cite{67}, the authors proposed a robust precoding scheme considering the unknown radar cross section variances and delays.
Qian \emph{et al.} paid attention to the effect of multiple clutters and proposed two spectrum sharing schemes, which  maximizes the radar SINR and user rate, respectively.
Considering the imperfection of the channel state information (CSI), the authors pointed out that the proposed scheme lacks robustness, and perhaps the optimality should be given up for better resilience \cite{18}.  
Recently, Qian \emph{et al.} further studied a multi-target scenario and performed a joint communication code book and radar precoding design. The effective interference of the radar was minimized in this study \cite{112}.
Similar to \cite{4}, Biswas \emph{et al.} considered the coexistence of an FD communication system and a radar. Instead of the joint design, two precoding schemes were devised for communication and sensing, respectively. Practical issues of imperfect CSI and hardware distortions were addressed. In this study, users are also equipped with multiple antennas to further reduce the received interference  \cite{11}. However, the corresponding algorithm suffers from intractable complexity. More than $e^{10}$ complex multiplications are required in each optimization iteration, only for a two-UL-users and two-DL-users setting. In \cite{12} and \cite{13}, Li \emph{et al.} investigated the spectrum sharing between a matrix completion (MC)-based MIMO radar and the MIMO communication transmitter. The MC-based radar refers to the radar that only forwards sub-sampled data to the processing center, and  the original matrix is reconstructed by the MC.
This sub-sampling process changes the interfering space from the BS to the radar.  A larger null space is thus available for BS precoding. 
He \emph{et al.} paid particular attention to the radar waveform in the joint optimization design. Practical constraints of waveform similarity  and constant modulus were taken into account. The constant modulus constraint ensures a  low peak-to-average power ratio (PAPR) and a low sidelobe of the optimized waveform. But this constraint also brings heavy computing overheads \cite{17}. Instead of using the SINR as the radar metric, Cheng \emph{et al.} proposed a robust scheme to minimize the Cramér–Rao bound (CRB) of the direction-of-arrival angle. The user SINR and waveform similarity were considered in the constraints.
The alternating direction method of multipliers (ADMM) and SDP were applied for optimizing the radar and BS precoding \cite{108}.
Considering both inter-user interference and inter-system interference,  Rihan \emph{et al.} and Hong \emph{et al.} investigated the scheme of the interference alignment, the core idea of which is to align different interference vectors into a small space at receivers. In this way, the remaining complementary space is interference-free for data transmissions and target detection  \cite{14}\cite{15}. In addition, Crossi \emph{et al.} optimized the communication transmit variance, radar transmit power, and radar linear filters to maximize the EE of the communication system while satisfying a predefined SINR of the radar in all range resolution bins \cite{75}. 

\subsubsection{Discussion}
We summarize these studies in Table \ref{TABLE1} by comparing their settings, optimization variables, programming algorithms, etc. It is obvious that MIMO mainly undertakes the role of spatial separation to make the two functions not interfere with each other. Transmit beamforming was the most addressed issue. It directs communication and sensing signals into spatially separable spaces at receivers. Multi-antenna receivers support spatial filters to further exclude residue
interference. As for sensing functions, MIMO support the phase and amplitude adjustment to shape the waveform. However, the network suffers from heavy burdens due to the high dimension of MIMO. The overheads of channel estimation
and feedback scale linearly with the antenna numbers. Much more power is required to use the large array and perform  multi-stream processing. The processing complexity scales polynomially and even exponentially with antenna numbers.

We may find that current JCAS MIMO schemes need communication and sensing systems tightly cooperated \cite{90}. They assumed the timely CSI is available and that two systems could constantly make adaptations to CSI changes. 
In addition, the same sampling rate, same symbol duration, and accurate time, frequency, and phase synchronization are also required for most schemes. The network is not robust for that any out-of-sync would greatly impair system performance.
Therefore, we may think the loose cooperation is more friendly in the practical implementation. In detail, we use the large-scale CSI rather than the full CSI to coordinate the two systems. The large-scale CSI is slowly varying and highly depends on the surroundings. Thanks to these properties, it could be predicted rather than using pilots for  estimation. To do so, a new architecture named the radio map could be introduced \cite{104}. It collects position information and maps the position data into the large-scale CSI.
On this basis, the large-scale CSI is used to instruct the coexistence design. It has been shown that this scheme is effective to decouple the interference geographically \cite{102-1,103,105}. By the way, we must admit that using the large-scale CSI inevitably leads to performance loss, but in turn, it costs less. The cost effectiveness ratio is of great importance for the large-scale deployment in reality.

\subsection{Communication and Sensing Integration}
Unified hardware leaves out many troublesome problems compared with separate configurations. The troubling issues of signaling, information exchange and synchronization are bypassed. In the literature, we use the term ``integration” to depict this relationship of communication and sensing. To endow one platform with two functions, the first mission is to balance the communication and sensing requirements in the waveform design. In the literature, there are three main ideas for the unified configuration. The first tries to adjust radar platforms for communication. Different information embedding schemes are proposed to let radar waveforms carry  information. The second attempts to assign sensing tasks to the communication system. Different estimation and detection
schemes are evaluated for  communication signals.  The last gives efforts to the dual-function waveform design, which achieves the most balanced communication and sensing performance.

\subsubsection{Radar-Centric Designs}
\begin{table*}[t]
	\tablefootnote{$f_{PRF}$ is the pulse repetition rate, $N_{sym}$ is the bits carried by one symbol, $M_r$ is the radar antenna number, $K$ is selected transmit number, $Q$ is the hop number of one pulse, $f_{SW}$ is the sweep frequency, $M_c$ is the code number.}
	\centering
	\caption{MIMO Enabled Radar-Centric Design for communication and sensing integration}
	\label{table1}
	\begin{threeparttable}
	\begin{tabular}{|p{2.3cm}|p{2.3cm}|p{2.5cm}|p{4.5cm}|p{3.5cm}|}
		\hline
		\textbf{Radar Waveform}&\textbf{Embedding Scheme}&\textbf{Directional/Broadcast Communication}&\textbf{Communication Rate}\tnote{1}&\textbf{Impacts on Sensing}\\ 
		\hline
	    \multirow{5}{*}{optimized waveform}
	   &AM&directional&$f_{PRF}*N_{sym}*M_r$&sidelobe variations\cite{23}\\
	    \cline{2-5}
	    &PM&directional, broadcast &$f_{PRF}*N_{sym}$& RSM\cite{24}\\
	    \cline{2-5}
	   	&PM&directional, broadcast &$f_{PRF}*N_{sym}*M_r$& RSM\cite{37,46}\\ 
	    \cline{2-5}
	    &PM\&AM&directional&$f_{PRF}*N_{sym}$& sidelobe variations\cite{25}\\
	    \hline
	    frequency diverse&FM&directional &$f_{PRF}*N_{sym}*(M_r-1)$& unspecific\cite{41}\\
	    \hline
	    unspecified&IM&directional&$f_{PRF}*N_{sym}*\lfloor \log(M_r!) \rfloor$&RSM\cite{26,27}\\
	    \hline
	    optimized waveform&IM&directional&$f_{PRF}*\lfloor \log(K!*C_{M_r}^{K}) \rfloor$& peak ripples of the main beam\cite{28}\\
	    \hline
	    FH&IM&broadcast&$f_{PRF}*Q*\lfloor \log(C_{M_c}^{M_r}) \rfloor$&RSM \cite{33}\\
	    \hline
	    FH&PSK,DPSK,CPM&broadcast&$f_{PRF}*Q*M_r*N_{sym} $&spectrum widening, RSM\cite{30,31,32}\\
	    \hline
	    FH&FSK&broadcast&$f_{PRF}*Q*M_r*N_{sym} $&frequency shift\cite{34}\\
	    \hline
	    LFM&PSK&broadcast&$f_{PRF}*N_{sym}$ or $f_{PRF}*M_r*N_{sym}$&grating lobe of the AF\cite{36}\\
	    \hline
	    PMCW&DPSK&broadcast&$f_{SW}*N_{sym}$&unspecific\cite{44} \\
	    \hline
	    FMCW&IM\&PM&broadcast&$f_{SW}*N_{sym}*\lfloor \log(K!*C_{M_r}^{K}*C_{M_f}^{K}) \rfloor$&unspecific\cite{45}\\
	    \hline
	\end{tabular}
	\begin{tablenotes}    
	\footnotesize               
	\item[1] $f_{PRF}$ is the pulse repetition frequency, $N_{sym}$ is the bits carried by one symbol, $M_r$ is the number of radar transmit antennas, $K$ is the number of selected antennas for one transmission, $Q$ is the hop number of one pulse, $f_{SW}$ is the sweep frequency, $M_c$ is the code number and $M_f$ is the carrier frequency number.           
	\end{tablenotes}
	\end{threeparttable}    
	\label{table_MAP}
\end{table*}

Using radars for  data transmissions was proposed early for communication and sensing integration because the communication functionality could leverage the high power and large antenna arrays of radars. In the literature, how to embed information into radar waveforms without impairing the sensing function is the main investigated issue. 
An early idea is to embed information in the radar sidelobe so that the mainlobe is unchanged to ensure the sensing performance. 
In  \cite{23}, the authors used amplitude modulation (AM) to modulate the sidelobe to convey information. The waveform diversity is applied to transmit multiple symbols in parallel.
However, this scheme has a fatal drawback of being powerless to serve the user in the mainlobe. Motivated by this, Hassanien \emph{et al.} further proposed  phase-modulation-based solutions \cite{24,46,37}. The phase modulation (PM) was implemented by a bank of weighted vectors, each of which constructs a certain phase offset. Since only the phase is altered to convey information, the information can thus be embedded in both the mainlobe and the sidelobe. Moving forward, Ferreira \emph{et al.} proposed a robust dual modulation scheme in which the PM and AM were selectively applied according to the user in the mainlobe or not. In this study, both  PM and AM vectors are expressed in closed forms, which not only gets rid of the troubling optimization but also improves the robustness to combat angular errors in case the radar does not know the exact user position \cite{25}. Moreover, Ji \emph{et al.} proposed an embedding scheme to map the binary bits into  
positive or negative frequency increments using the frequency-diverse MIMO radar. This kind of radar is characterized by high resolution because its waveform is  range- and angle-dependent \cite{41}. By realizing this, Alselwi \emph{et al.} divided the transmit platform into overlapped subarrays and assigned incremental frequencies to these subarrays. A closed-loop architecture was devised, that uses the detection feedback as the indicator to choose transmit beamforming vectors and embedding schemes. The phase shift key (PSK)  and amplitude and phase shift key (APSK) were applied to for the user in the mainlobe or not \cite{56}.


In addition to traditional modulation schemes, the index modulation (IM) was investigated in \cite{26,27,28,33}, which uses the index of a candidate set to convey information. When the candidate set is the waveform,  information is embedded by shuffling  different waveforms across transmit antennas. This embedding scheme is transparent to radar operations for that the matched filtering at radar receivers undoes this shuffling, enabling the radar operation to proceed as if it is a radar-only platform \cite{26,27}. Wang \emph{et al.} further applied a sparse array and combined antenna selections and permutations to achieve the IM \cite{28}. When the candidate set turns to the transmit codes,  information is thus embedded by pairing the codes with transmit antennas. In \cite{33}, the frequency-hopping (FH) radar was investigated to carry information by assigning different FH codes to different antennas. Under this design, the  waveform has a better spectrum profile than the PSK-based scheme, because the phase discontinuity of the latter brings significant power variations among different spectra.  In short, thanks to multiple permutations and combinations, the communication rate under IM could achieve several Mbps. 

In addition to \cite{33}, FH radars have also been investigated in \cite{30,38,34,31,32}. This kind of radar divides one pulse into multiple sub-pulses, namely, hops, and uses different frequencies in different hops. Information bits are thus  embedded in each hop rather than each pulse so that the communication rate is improved. The corresponding schemes of PSK \cite{30}\cite{38}, frequency shift keying (FSK) \cite{34}, differential phase shift keying (DPSK) \cite{31},   and continuous phase
modulation (CPM) \cite{32} were successively proposed. Similarly, the linear frequency modulation (LFM) waveform was applied in \cite{36}, where the sweep time was divided into multiple subunits for information embedding.   However, these fast-time embedding schemes are sensitive to inaccurate factors. Corresponding issues of time offset \cite{19}, channel estimations \cite{20}, and engineering-friendly coding and decoding schemes have not been fully investigated \cite{21}. 
From the perspective of sensing, the embedding impacts on the sensing function should be evaluated. Analysis found that after embedding, the range sidelobe is reduced due to the randomness of embedded symbols, which lowers the correlation among radar pulses \cite{109}. However, such randomness also causes pulse variations within the coherent processing interval and leads to the range sidelobe modulation (RSM), which weakens the target visibility \cite{39}. 

\begin{figure*}[!htbp]
	\centering	
	\subfigure[Minimal sidelobe versus user number]
	{
		\includegraphics[width=0.45\textwidth]{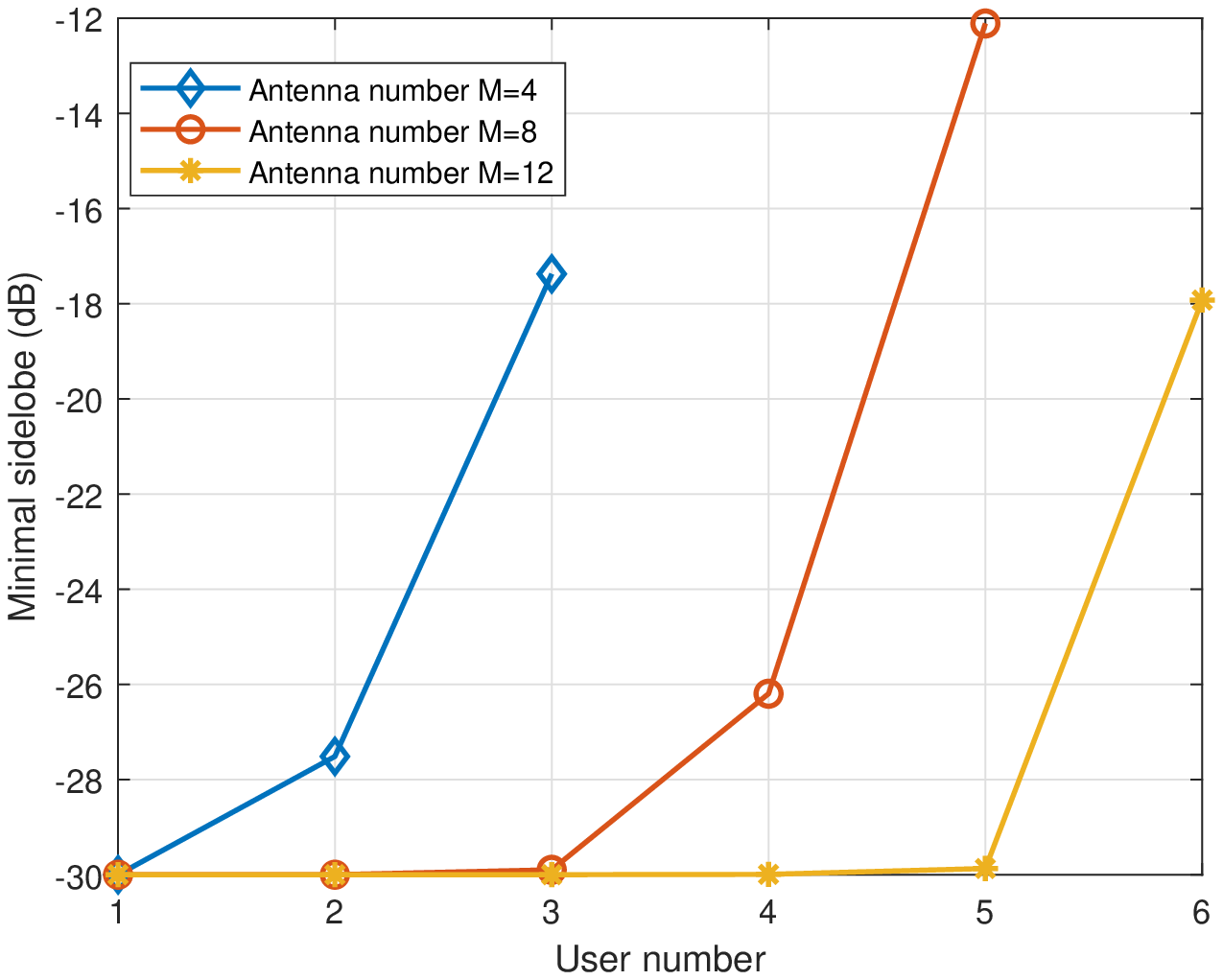}
		\label{a}
	}
	\subfigure[Transmit beampattern]
	{
		\includegraphics[width=0.45\textwidth]{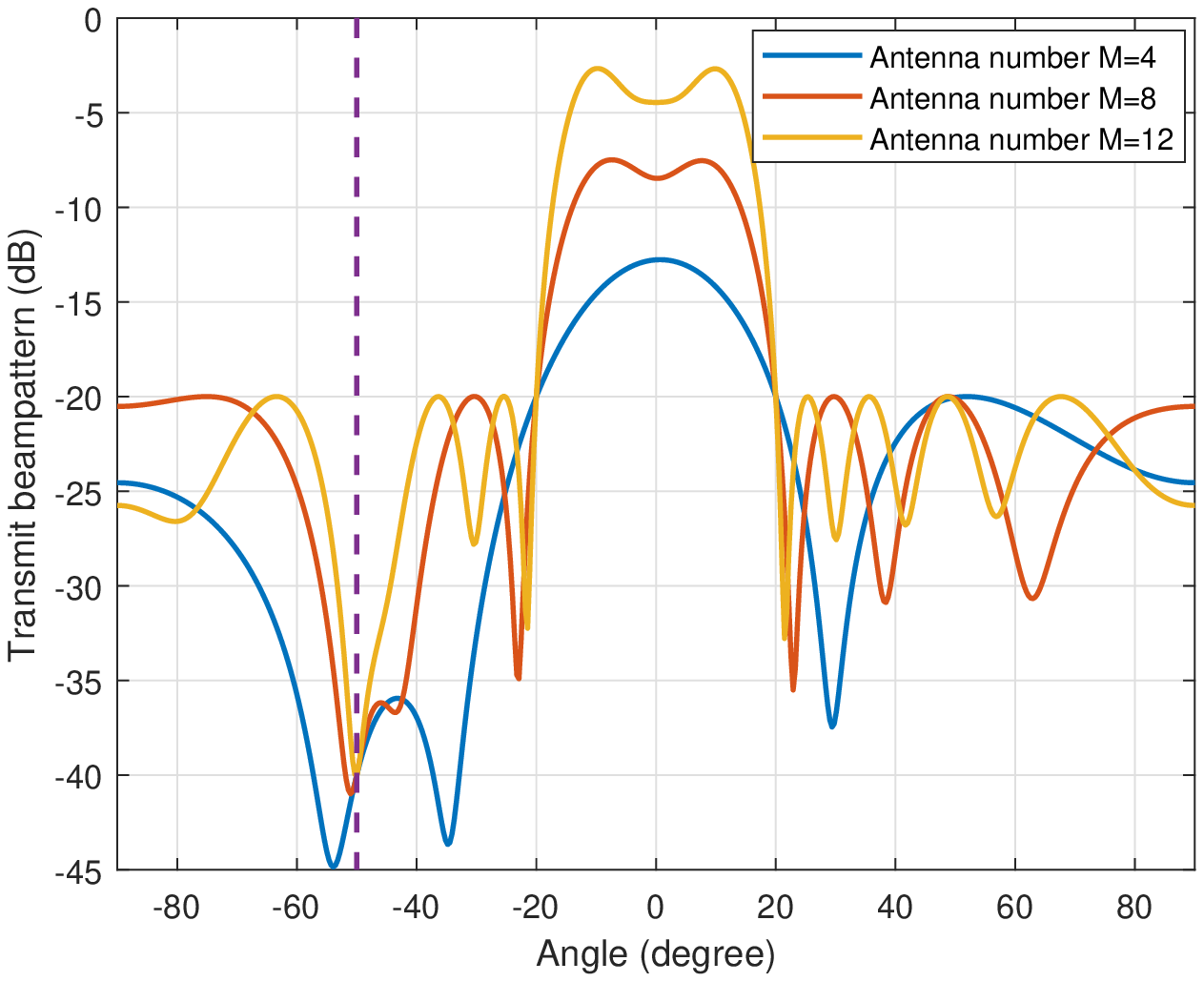}
		\label{b}
	}
	\caption{Illustration of the trade-off between communication and sensing and the transmit beam pattern under different transmit antenna numbers. The figure is obtained under the following setting. We assume the radar mainlobe is in the direction of $[-10^{\circ},10^{\circ}]$ and the desired waveform is set as $e^{-j\pi\sin(\theta)}$, where $\theta$ denotes the spatial angle. The sidelobe power is assumed to be $20$ dB lower than the mainlobe, i.e., $\epsilon=0.1$. One radar waveform is assumed to carry one bit. $\delta^2=0.001$ represents transmitting symbol 1 and $\delta^2=0.0001$ represents transmitting symbol 0. The six users are located at the direction of $[-30^{\circ},-40^{\circ},-50^{\circ},-60^{\circ},-70^{\circ},-80^{\circ}]$, respectively. 
	The right subfigure considers one user who is located in the direction of $-50^{\circ}$ and the symbol $0$ is transmitted.}
	\label{adfig1}
\end{figure*} 

The above embedding schemes all use pulse radars as the dual-function radar and communication (DFRC) platforms. Another kind of radar, named the continuous wave radar, has also been evaluated. Different from pulse radars,  continuous wave radars continuously emit probing signals and  receive echoes. To reduce the signal leakage from transmitters to receivers, the continuous wave radar usually uses low-power signals and small antenna arrays and has a small size. These traits make them  friendly to civil applications.
To ensure the accuracy of the parameter estimation, Dokhanchi \emph{et al.}  applied a multi-antenna frequency modulated continuous waveform (FMCW) radar and proposed a DPSK modulation scheme \cite{44}. For vehicular applications,  Ma \emph{et al.} investigated a hybrid PM and IM scheme operating over the phase modulated continuous waveform (PMCW), which achieves better a bit error rate (BER)  than the PM-only modulation. In this design, the complexity and cost are well controlled by using the sparse array and reduced RF modules \cite{45}.

To show the potential trade-off between communication and sensing and demonstrate the advantage of using multiple antennas, we apply the scheme proposed in \cite{23} as an example. In detail, the corresponding scheme optimized the transmit waveform to minimize its difference from a desired waveform in the mainlobe and limits the sidelobe within a predefined threshold, i.e., $\epsilon$. To convey information in the sidelobe, there is another constraint that forces the radiating power toward user directions to be equal to a certain value, i.e., $\delta^2$. This value is chosen from a predefined set and is determined by the transmit symbol. Although the sidelobe threshold $\epsilon$ is given in advance, there is a minimal value to guarantee the problem feasibility. In Fig. \ref{a}, we calculate the minimal sidelobe power, i.e., $\epsilon^2$, under the situation of  different number of users.  It can be seen that the sidelobe increases with the number of users. Since the high sidelobe increases cluttering effects in the AF and leads to target invisibility, serving more users is at the expense of sensing accuracy. Fig. \ref{b} shows the superiority of using multiple antennas. The radar radiating signal shows a more desirable pattern when more transmit antennas are available.

\emph{Summary:} We summarize the radar-centric schemes in Table \ref{table_MAP} by comparing their embedding methods, communication rate, the impacts on sensing, etc. 
Since communication is the secondary function in these designs, it suffers from the low rate, high outage probability, and weak serving ability, e.g., only supporting one user at a time. In these designs, the MIMO configuration improves the compatibility with the communication function. Supported by the waveform diversity, MIMO radar could simultaneously bear multiple symbols using the orthogonal waveform and improve the communication rate to the Mbps level. Controlled by multiple RF chains, radar beams could flexibly
carry different information toward different directions and concurrently serve several users. As for the primary sensing function, MIMO improves its resistance to the waveform distortion. However, despite using MIMO, it is still far from satisfying communication and sensing performance. First, information embedding brings more or less alterations to the radar waveform, most of which are negative. The non-constant envelopes, widening spectrum, RSMs, and grating ambiguity function (AF) patterns are all possible negative impacts \cite{23,93,94}. 
Second, the communication performance is far from that of cellular systems.  As for the most investigated DL information embedding, practical issues of inaccurate CSI,  time/frequency/phase offset and  coding and decoding schemes have  not been fully addressed \cite{21,94}. In terms of UL access, new mechanisms should be introduced to respond to  random user requests.  The most challenging issue is the differentiation of target echoes and user signals. Note that the target echo is likely submerged in noises for it experiences two-fold attenuation. If users and targets are located in different directions, directional T\&R beamforming works to divide communication and sensing signals, but when  users and targets are rightly in the same resolution bin, separating the mixed signals is not an easy task \cite{94}.

\subsubsection{Communication-Centric Designs}

\begin{table*}[t]
	\caption{MIMO Enabled Communication-Centric Design for Communication and Sensing Integration}
	\label{table4}
	\begin{tabular}{|p{2.5cm}|p{2.5cm}|p{3.2cm}|p{4.3cm}|p{3.2cm}|}
		\hline
		
		\textbf{Communication Waveform}&\textbf{Sensing Setting}&\textbf{Sensing Signal}&\textbf{ Sensing Performance}&\textbf{Required Change}\\ 
		\cline{1-5}
		single-carrier waveform&T\&R collocated&IEEE 802.11ad frame&cm-level range resolution, cm/s-level velocity resolution using 1.76GHz bandwidth&FD designs \cite{70} \\
		\cline{1-5}
		OFDM&T\&R collocated&5G new radio signals&tens of targets, sub-meter level location resolution at 28GHz&FD designs \cite{n4} \\
		\hline
		OFDM&T\&R separate, passive sensing&preamble of IEEE 802.11a frame&Improved detection performance compared with blind estimation&not required \cite{71}\\
		\hline
		unspecific&T\&R separate&demodulation reference signal&0.1m location error of the bi-static setting using 400MHz bandwidth&time synchronization\cite{115}\\
		\hline
		OFDM&T\&R collocated&varied version of the OFDM frame&15m range resolution of 0.25MHz subcarrier bandwidth&FD designs \cite{72} \\
		\hline

		OFDM&T\&R separate&the whole frame&unspecific&An additional separate receiving antenna \cite{113}\\
		\hline
	
	\end{tabular}
\end{table*} 
Using communication signals for sensing is another way to integrate these two functions. 
We  put our focus on DL sensing because BSs are more likely to be equipped with massive antennas and have strong signal processing ability.  
As for DL sensing, the most straightforward way is to directly use communication signals and extract the target information from their echoes. 
In \cite{70}, the preamble of the single-carrier frame of IEEE 802.11ad was evaluated to conduct target detection and parameter estimation. Thanks to the perfect auto-correlation property of the Golay complementary sequences in the preamble, it achieves cm-level range accuracy and cm/s-level velocity accuracy.
In \cite{123}, Kumari \emph{et al.} devised a sparse analog beamforming scheme for IEEE 802.11ad-based waveform. In each transmission, a subarray is randomly activated, and the receiver obtains the target information from the randomness-incurred grating lobes. In \cite{n4}, Pucci \emph{et al.} evaluated the sensing performance of the 5G new radio. Instead of directly using communication signals, communication and sensing beams are spatially separated, and the power of orthogonal frequency division multiplexing (OFDM) signals is split for these two functions. Through a series of evaluations, the results show that tens of targets could be detected with sub-meter level accuracy. This verifies the feasibility of endowing the communication system with sensing functions. 
In \cite{71}, the OFDM physical layer convergence protocol preamble of IEEE 802.11a was used for passive sensing. Passive sensing refers to that there is no cooperation between communication transmitters and sensing receivers. refers to there is no cooperation between communication transmitters and sensing receivers. By exploiting the prior knowledge of the preamble, the detection performance was largely improved compared with the blind estimation. Using the 5G new radio demodulation reference signal,  Kanhere \emph{et al.} estimated location parameters under the bi-/multi-static configurations. The estimation error was analyzed under different geometrical T\&R settings \cite{115}.
Rahman \emph{et al.} developed a C-RAN-based OFDM MIMO perceptive network. The authors clarified three sensing scenarios and investigated two parameter estimation methods. A clutter reduction scheme was also proposed to improve the sensing performance \cite{92}.

Directly using standard communication signals for sensing is not the optimal choice. 
Due to the randomness of carried symbols, communication echoes present high-range sidelobes after correlation-based processing. As a result, some adaptations are introduced to current communication systems. Liu \emph{et al.} used a modified OFDM waveform and devised a high-resolution range estimation algorithm by effectively using the whole array aperture and the available bandwidth \cite{72}. System parameters to meet both communication and sensing requirements were further discussed in \cite{73}. In \cite{113}, Ni \emph{et al.} added a separate receiving antenna to BSs to bypass the FD problem. A waveform optimization scheme was proposed to achieve balanced communication and sensing performance. 

\begin{figure}[tbp]
 	\centering	
 	\includegraphics[width=0.5\textwidth]{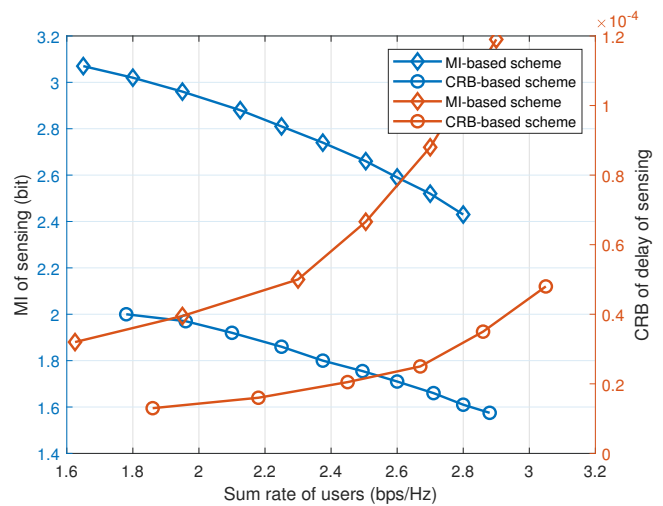}
 	\caption{Illustration of the trade-off between IM or CRB of sensing and  sum rate of users under the scheme proposed  in \cite{113}. The setting includes two users, three targets and one BS with 16 antennas. Other parameters are omitted here and readers can refer to \cite{113} for more detailed information.}
 	\label{adfig2}
 \end{figure}

To show the potential trade-off of communication and sensing, we take the scheme proposed in \cite{113} as an example. In \cite{113}, the authors maximized communication SINR under the radar constraint of mutual information (MI) or CRB.  This constraint requires the MI to be greater than a predefined threshold or limits the Euclidean distance between the actual precoding vector and the optimal CRB precoding vector. As shown in Fig. \ref{adfig2}, we depict the relationship between MI or CRB  and the sum rate of users. It can be seen that the user rate has to make a certain compromise if a good sensing performance is required. In turn, if a high data-rate transmission shall be satisfied, there is a great loss of the sensing accuracy. This proves that the interests of communication and sensing is somewhat conflicting.


\emph{Summary:} We summarize these studies in Table \ref{table4} by comparing their sensing settings,  signals, and performance. In these designs, MIMO brings extended spatial ``receptive filed" to the communication transceivers to sense the surroundings.  In  current 5G networks, cellular BSs are densely deployed, and these infrastructures offer ubiquitous platforms for sensing. If corresponding schemes become mature, the communication network that works as data pipes would upgrade to the perceptive network and better support massive applications. 
However, before practical deployments, several technical obstacles must be overcome.
First, the impact of imperfect CSI and out-of-sync should be fully evaluated. As for the OFDM waveform, it is sensitive to the carrier offset. The inter-carrier interference would largely degrade the sensing accuracy \cite{74}.
The second one is the FD problem. If  BSs upgrade to dual-function platforms, the original UL\&DL separate working regime is not applicable. They have to simultaneously transmit signals and receive echoes. 
Effective schemes to isolate the signal leakage from transmitters to receivers are still lacking \cite{83}. Similar to radar-centric designs, the differentiation of echoes and user signals is not an easy task. More advanced signal processing is needed.
Promisingly, the UL user signal also carries target information and can be further utilized \cite{120}.
Another direction is bi-static and passive sensing, in which sensing receivers are not collocated with  transmitters.
There are two main problems: interference and synchronization. If T\&R do not cooperate, the received signal is more likely to be a mixed one of many signals  \cite{83}. It is difficult to extract information from this noise-like mixture. 
However, even if a clear signal is obtained, imperfect synchronization can lead to sensing ambiguity due to the failure of the preamble location \cite{82}. Perfect time/frequency/phase synchronization is a challenging issue if there is no wired link to connect T\&R devices. 

\subsubsection[Novel Waveform Designs]{Novel Waveform Designs}
Designing a novel waveform moves a further step for communication and sensing integration. In the literature, there are three different hardware configurations. The first uses separate communication and sensing modules to bypass their incompatibility. In this situation, the mutual interference was the most addressed topic.
Temiz \emph{et al.} applied a novel communication transmit precoder that effectively exploits  radar interference. The radar signal was optimized to approximate the communication signal so as to reduce the precoding cost of eliminating radar interference \cite{54}. On this basis, two optimal power allocation schemes were devised in \cite{55}, with the objective of  sum rate and energy efficiency. Recently, the authors further detailed the UL receiving procedure, where the user messages and target echoes are successively processed using the SIC. Analytical works were further conducted under the consideration of imperfect CSI and self-interference, and the inner trade-off  between the achievable rate and radar detection performance was revealed \cite{110}.
Liu \emph{et al.} proposed two precoding schemes to compare the performance of  separate and shared deployments. The former divides antennas into two groups, and they are connected to independent communication and sensing modules,  while the latter uses unified modules. The results show that the unified configuration achieves better performance due to more available DoFs \cite{50}. 
In \cite{63}, Dong \emph{et al.} investigated the same separate setting as \cite{50} and  proposed a new algorithm to reduce the problem-solving complexity. 

\begin{table*}[t]
\begin{threeparttable}	
	\centering
	\caption{MIMO Enabled Novel Waveform Design for Communication and Sensing Integration}
	\label{table3}
	\begin{tabular}{|p{2cm}|p{2.5cm}|p{3cm}|p{2.5cm}|p{3cm}|p{3cm}|}
		\hline		
		\textbf{Integration Degree}&\textbf{Setting}\tnote{1}&\textbf{Optimization Variable}&\textbf{Sensing Metric}&\textbf{Communication Metric}&\textbf{Programming Scheme}\\ 
		\hline
		\multirow{4}{1.9cm}{separate RF chains and separate antennas}&MT-MIMO,MU-MISO&radar waveform and power, BS transmit precoding and power&waveform similarity&sum rate and EE &unspecific\cite{55} \\
		\cline{2-6}
		&MT-MIMO, MU-MISO&radar transmit covariance matrix, BS precoding&waveform similarity&SINR &SDP \cite{50}\\
		\hline
		\multirow{6}{1.9cm}{separate RF chains and shared antennas}&
		ST-MIMO, SU-MISO&joint radar and BS precoding&desired radar signal, constant envelop& desired communication signal&error reduction algorithm \cite{59}\\
		\cline{2-6}
		&MT-MIMI,MU-MISO&hybrid beamforming&waveform similarity&SINR&second order cone programming \cite{n2}\\
		\cline{2-6}
		&ST-MTMO, SU-MIMO&radar transmit signals and precoding, communication transmit covariance&SINR&secure rate&SDP\cite{66}\\
		\cline{2-6}
		&Multiple ST-MIMO, MU-MIMO&transmit covariance matrix of radar and BS, computation resource& desired radar pattern&transmission delay&fractional programming, Taylor expansion \cite{n1} \\
		\hline
		
		\multirow{15}{1.9cm}{share all the components}&ST-MIMO, MU-MISO&waveform matrix&range sidelobe and waveform similarity&multi-user interference&Riemannian gradient conjugate algorithm\cite{58}\\
		\cline{2-6}
		&MT-MIMO, MU-MISO&waveform matrix&waveform similarity&multi-user interference&Riemannian gradient conjugate algorithm\cite{162}\\
		\cline{2-6}
		&ST-MIMO, MU-MISO&transmit baseband and RF precoding, receive RF precoding& radar power &gain of each user&unspecific \cite{121}\\
		\cline{2-6} 
		&MT-MIMO, SU-MIMO&transmit covariance matrix &CRLB of locations&capacity&SDP and sequential parametric convex approximation\cite{61}\\
		\cline{2-6}
		&ST-MIMO, SU-MISO&joint T\&R precoding&Kullback-Leibler divergence, PAPR&word error probability& alternating direction sequential relaxation programming\cite{64} \\
		\cline{2-6}
		&MT-MIMO, MU-MISO&joint radar and BS precoding&waveform similarity, cross correlation&SINR&QSQP \cite{51}\\
		\cline{2-6}
		&ST-MIMO, MU-MISO&joint precoding&CRB&SINR&SDP \cite{111}\\
		\cline{2-6}
		&ST-MIMI, MU-MISO&transmit and receive vectors&SINR, PAPR and constant modulus&constructive interference instructed constructive region requirement&majorization-minimization, ADMM \cite{n3} \\
		\cline{2-6}
		&MT-MIMO, MU-MIMO&transmit and receive vectors&a broad family of radar metrics&user error rate& alternating optimization \cite{n8} \\
		\cline{2-6}
		&MT-MIMO, MU-MIMO&transmit vector&effective receiving power and cross correlation of targets&sum rate&successive convex approximation \cite{n9}\\
		\cline{2-6}
		&ST-MIMO, SU-MISO& transmit signal and distortion &SINR&secrecy rate& Taylor series approximation\cite{52}\\
		\cline{2-6}
		&ST-MIMO, MU-MIMO&transmit signal and artificial covariance matrix &waveform similarity&SINR&fractional programming\cite{53}\\
		\hline
	\end{tabular}
	\label{TABLE3}

\begin{tablenotes}
	\footnotesize
	\item[1] ST-MIMO: single-target MIMO, MT-MIMO: multi-target MIMO, SU-MISO: signle-user multi-input and single-output (MISO), MU-MISO: multi-user MISO.
\end{tablenotes}
\end{threeparttable}
\label{TABLE}
\end{table*}
The second configuration uses separate RF chains but the same antenna array to transmit or receive communication and sensing signals. As for this configuration, the combined waveform of communication and sensing would impinge the targets and determine the sensing performance.  McCormick \emph{et al.} investigated a spatial beamforming scheme that considers the high PAPR problem of the formed waveform. In addition, the constant-envelope constraint was considered in the beamforming design \cite{59}. In \cite{60}, the authors further conducted practical experiments to verify the effectiveness of the proposed beamforming scheme. Similarly, Jiang \emph{et al.} used linear superposition to combine spatially separate communication and sensing beams. The constant-modulus constraint was considered to overcome the high PAPR problem of the formed waveform \cite{117}. Liu \emph{et al.} put the cross correlation into account and proposed a joint communication and sensing precoding scheme. The waveform was optimized to approach the desired one  \cite{51}. 
Buzzi \emph{et al.} investigated the configuration of mMIMO and applied zero-forcing beamforming for communication signals and channel matched beamforming for sensing signals  \cite{122}. Qi \emph{et al.} proposed a hybrid  beamforming scheme to detect multiple targets and communicate with multiple users at the same time. The transmit pattern was optimized under the constraint of the user SINR \cite{n2}. 
In reality, the detected targets may be malicious eavesdroppers, especially in the military. 
By taking this into account, Chalise \emph{et al.} used the secure rate as the communication metric and put up a joint optimization scheme, including the radar waveform, radar precoding, and the communication transmit covariance \cite{66}. 
Moreover, Ding \emph{et al.} jointly considered the MIMO design with the computation resource allocation. In the proposed network, multiple dual-function user terminals perform communication and sensing tasks simultaneously. Their sensing data are processed locally or offloaded to the mobile edge computing  (MEC) sever. To obtain good sensing performance and limit the offloading time, the authors jointly optimized communication and radar precoding matrices and computation resources. This work did a pioneer study for the combination of sensing, communication and computing \cite{n1}.

The third configuration uses totally unified modules to exploit the ultimate utilization of the hardware, space and energy.  In \cite{57}, Liu \emph{et al.} derived the optimal expressions of  omnidirectional and directional sensing patterns. On this basis, the authors proposed a weighted optimization scheme that achieves a flexible trade-off of communication and sensing. Then, the authors further took the sidelobe control into account. An improved weighted objective that includes the sidelobe level, waveform similarity and user interference was optimized \cite{58}. To utilize the inter-user interference for symbol discrimination, Zhang \emph{et al.} devised two constructive-interference-based waveform designs \cite{162}. Compared with the scheme proposed in \cite{57}, this scheme achieves a lower symbol error rate.
Liyanaarachchi \emph{et al.} used hybrid transmit beamforming to generate dual-function beams and used null-space projection to mitigate the inter-user interference. The received beamforming was also devised to reduce the impact of self-interference \cite{121}. 
To exploit the high resolution of the frequency diverse waveform, Zhou \emph{et al.} investigated the phase-modulated frequency diverse waveform to perform both communication and sensing tasks. The sensing metric of the Cramér–Rao lower bound (CRLB) of locations and the communication metrics of the capacity and the BER were analytically derived, and the transmit precoding vector was optimized \cite{61}.  Recently, Wang \emph{et al.} investigated the waveform design for joint synthetic aperture radar imaging and communication. Two novel waveforms of modified OFDM and space-time coding  were devised to accomplish high-resolution imaging and data transmissions at the same time. On this basis, the authors pointed out that multidimensional waveform optimization is effective in mitigating the conflicts of communication and sensing \cite{62}.
Tian \emph{et al.} followed the idea of sidelobe control proposed in  \cite{23}  and refined this scheme by joint T\&R beamforming. Simulation results show that the new optimized waveform achieves better communication and sensing performance than directly embedding bits in the radar waveform \cite{64}. In particular, instead of using indirect objectives such as waveform similarity and radar SINR, Liu optimized the CRB to guarantee the estimation performance, which, however, leads to a much more complex problem \cite{111}. 
More recently, Liu \emph{et al.} jointly optimized the T\&R vectors by using the constructive-interference-based processing and  space-time adaptive processing. The proposed scheme is robust to the clutter environment \cite{n3}. 
Zhang \emph{et al.} devised a novel waveform that is characterized by the low sidelobe and ultra reliability. The main idea of this design is to map the information bits into a high-dimensional sparse vector. Thus, great spatial diversity can be obtained to ensure reliability \cite{n7}. Johnston \emph{et al.}  investigated the OFDM-DFRC system. The radiated waveform and receive filters were jointly optimized under the constraints of the user error rate and transmit beam patterns \cite{n8}. In particular, the authors optimized a general sensing objective, which can be transformed into any specific sensing metrics in practical applications.  
In addition, Wang \emph{et al.} applied the non-orthogonal multiple access (NOMA) to serve communication users. The  transmit precoding was optimized to maximize the weighted objective of the sum rate and the effective receiving power of sensing targets \cite{n9}.
Regarding physical security, Deligiannis \emph{et al.}  employed the artificial noise to control the received  SINR of a malicious target \cite{52}.  Su \emph{et al.} extended this study to a multi-user setting. A joint precoding and artificial noise optimization scheme was devised to minimize the target SINR while guaranteeing the SINR of legal users. This work also modeled the uncertainty of the target using the imperfect and statistical CSI \cite{53}. 

\begin{figure}[tb]
	\centering	
	\includegraphics[width=0.5\textwidth]{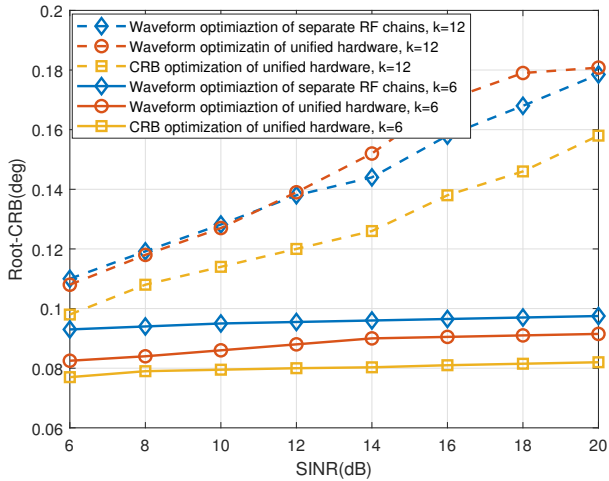}
	\caption{Illustration of the trade-off between root-CRB of sensing and communication SINR under the schemes proposed in \cite{50}, \cite{51} and \cite{111}. It is assumed that the DFRC platform is equipped with $16$ transmit antennas and $20$ receive antennas. The transmit power is $30$ dBw and the target is in the direction of $0^{\circ}$. Other parameters are omitted here for saving space. Readers can refer to \cite{111} for more detailed information. 
	\label{adfig3}}
\end{figure}

To show the overall performance of the novel waveform design. We compare the schemes proposed in \cite{50}, \cite{51} and \cite{111}. These three schemes all investigated a multi-user DFRC system. The scheme in \cite{51} is based on the second configuration, and schemes in \cite{50} and \cite{111} are based on the third configuration.
They all maximized the sensing performance and required the user SINR to exceed a predefined threshold. 
The difference is that schemes in \cite{50} and \cite{51} took the waveform similarity as the objective, while the scheme in \cite{111} took the CRB as the objective.  In Fig. \ref{adfig3}, we illustrate the root-CRB performance under these three schemes. It can be seen that the CRB-oriented optimization achieves the minimal root-CRB. This shows that using waveform similarity as the objective does not keep the complete consistency with real sensing metrics.   
In addition, we can see the trade-off of communication and sensing: increasing the user SINR increases the root-CRB.  But when the user number is small, i.e., $k=6$, the increasing trend of the root-CRB is not obvious. This is attributed to MIMO that brings surplus DoFs to relieve the conflicts of communication and sensing.  

\emph{Summary:}
We summarize these studies in Table \ref{TABLE3} by comparing their settings, schemes, and communication and sensing metrics.
Compared with radar-centric and communication-centric schemes, novel waveform designs are not restricted by the primary function so that full DoFs can be exploited. In terms of using one platform that includes totally separate communication and sensing modules, their generated beams are designed to be spatially separated. This is very similar to the communication and sensing coexistence. Since different RF chains and antenna arrays are applied, the separate configuration is able to achieve comparable performance as the communication and sensing standalone configuration while enjoying convenient access to communication and sensing information.
When communication and sensing signals are generated independently but are transmitted by the same antennas, considerable attention has been given to the transmit pattern of combined communication and sensing signals, whose properties directly determine the sensing performance. As for the communication function, spatial beamforming is widely applied to mitigate the interference from high-power sensing signals. The totally unified setting is the most cost-friendly, and the corresponding MIMO schemes mostly focus on the waveform pattern optimization and spatial interference mitigation. In these three settings, the communication and sensing trade-off mainly becomes the balance of waveform shaping and directional beamforming. The former guarantees the sensing performance, and the latter ensures the communication performance. With limited  DoFs, the objective of JCAS MIMO is to find a proper balance in different situations.

\subsubsection{Discussion}
At present, communication and sensing integration is still in its infancy. Making adaptations to current radars or BSs is the first step we take. Since the radar-centric and communication-centric schemes are 
constrained by the pre-given function, it is difficult to provide balanced communication and sensing services. 
Open issues in terms of FD operations, imperfect synchronization and CSI acquisition deserve further investigation. 
Moving a further step, devising a novel dual-function waveform is necessary for the long-term evolution of communication and sensing integration.
But the roles that MIMO undertakes, waveform shaping and directional beamforming, are not compatible. 
A theoretical breakthrough on the pareto optimal curve is expected for the multi-objective optimization.
The complexity problem also deserves more attention. Current waveform designs mainly turn out to be a high-dimension matrix optimization problem when using MIMO. 
In most cases, these non-convex problems are NP-hard. Even using the convex approximation to find a local optimum has  polynomial-level complexity. This  is far from engineering applications. 
To reduce the problem-solving complexity, one possible solution is to use intelligent methods. In the training stage,   a network is built by learning from the data. Then, in the inference stage, the trained network directly maps the input information into the output design.  Thus, the troublesome optimization is omitted in practical implementations. In particular, we emphasize the method of deep reinforcement learning (DRL). This method establishes a self-evolution network by constantly interacting with the environment. Based on this constant accumulation, the corresponding scheme is able to adapt to the environment in a progressive manner while maintaining small update complexity each time.


Furthermore, instead of using limited DoFs to chase the communication and sensing balance, which regards them as competitors, the interdependence of communication and sensing could be exploited. On the one hand, sensing results provide effective side information for communication. 
Sensing-assisted channel estimation \cite{125} and sensing-assisted beam domain designs \cite{81} have been investigated and proven to be  effective. From another perspective, communication also contributes to sensing. The sensing resolution could be refined by the signal-level and data-level fusion. If we take this reciprocity into account, the dilemma of using limited DoFs to cater to incompatible communication and sensing requirements may readily solved. This communication and sensing symbiosis may form a closed loop of communication-assisted sensing and sensing-assisted communication and finally achieve the ultimate performance of JCAS.

\section{Interplay of MIMO-Empowered JCAS and Cutting-Edge Technologies}
\label{sec}
In this section, we discuss three novel JCAS MIMO structures combined with cutting-edge technologies. 
The first is cooperative MIMO, which exploits not only  micro but also macro spatial DoFs.
The second introduces UAVs into the communication and sensing network. With terrestrial BSs, they cooperatively form a dynamic 3D MIMO structure. The last combines active MIMO and passive MIMO into the JCAS network. These MIMO structures bring new opportunities but also new problems for JCAS.

%
\subsection{JCAS with Cooperative MIMO}
\begin{figure}[htbp]
	\centering	
	\includegraphics[width=0.5\textwidth]{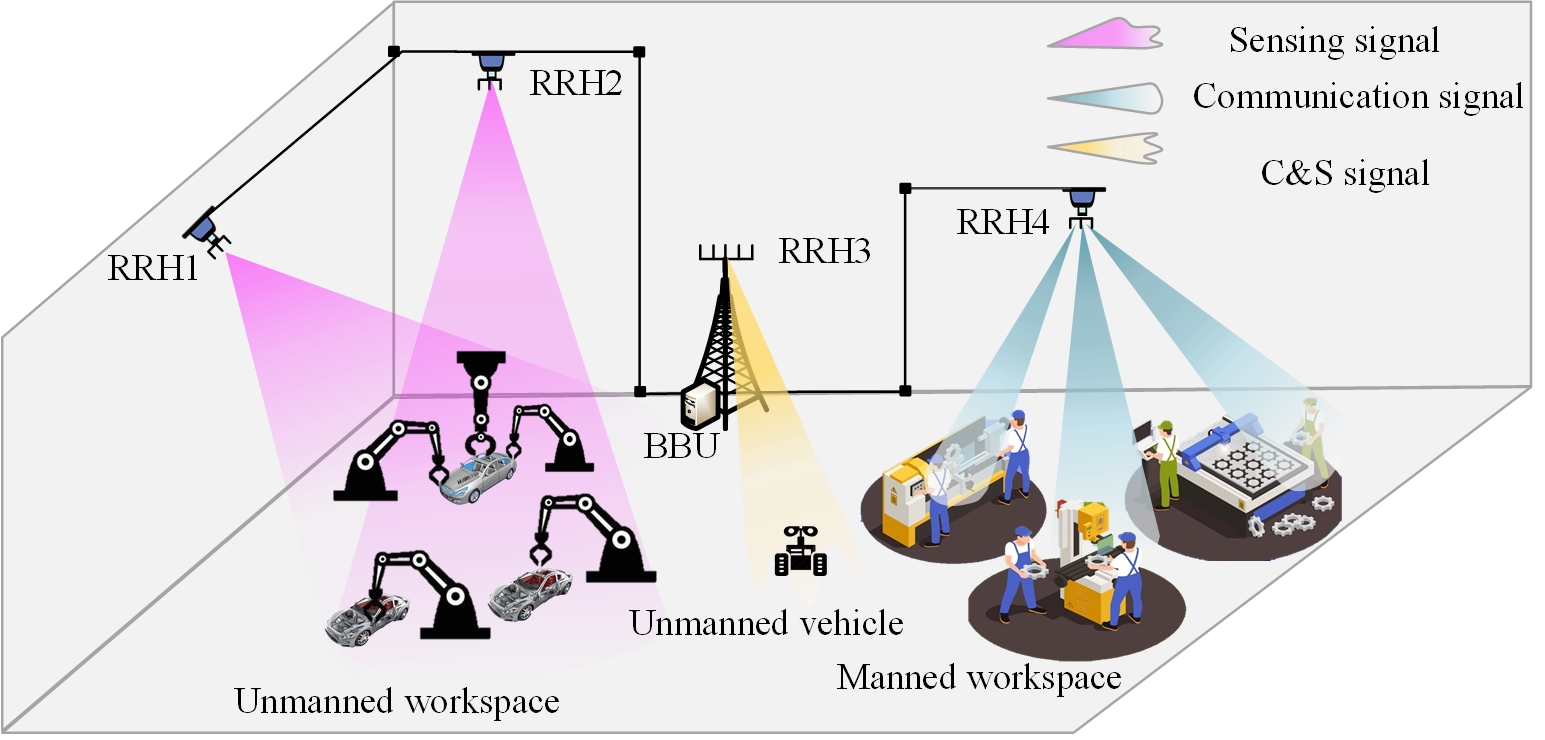}
	\caption{Illustration of a C-RAN architecture in the smart factory. RRHs are distributed in different places and transmit or receive different communication and sensing signals to serve different workers and machines. Signal processing and resource allocation are performed by the BBU. An MEC is maintained in the BBU to calculate the scheduling schemes and control the fabrication processes. The intelligent manufacturing is supported by the production-aware closed-loop optimization.}
	\label{fig2}
\end{figure} 


Cooperative MIMO refers to applying distributed nodes to transmit or receive signals, and these signals are jointly processed in a central unit. 
As for the communication side, a well-conditioned channel matrix could be constructed by selectively activating and muting different nodes at different times. Benefiting from multi-perspective observations, the sensing accuracy can also get great improvement. In addition,  we could assign different nodes with communication-only or sensing-only tasks so that the incapability of using one waveform for two uses is downplayed.
In short, the macro diversity not only improves individual communication and sensing performance but also gives the system more choices to configure the two functions in a more compatible manner.
In the literature, Ahmed \emph{et al.} considered a distributed DFRC system and proposed a power allocation scheme. The sensing performance is greatly improved by jointly processing the echoes from all DFRC nodes \cite{80}. Sanson \emph{et al.} considered a vehicle network and devised a cascading information fusion method to improve the resolution of the multi-target detection. Results show that, enhanced by the cooperative MIMO, the originally indistinguishable targets are distinguishable. The authors also conducted practical experiments to verify their results \cite{65}.

In 5G networks, an architecture named C-RAN is  a kind of cooperative MIMO. It consists of distributed remote radio heads (RRHs), a base  band unit (BBU) and a fronthaul network that connects RRHs to the BBU. The RRH is responsible for signal transmissions and receptions. Signal processing and resource allocation are performed by the BBU. 
This 5G architecture provides a ready-made platform to build the distributed perceptive network. 
In Fig. \ref{fig2}, we illustrate a smart factory that operates under the C-RAN. 
In the unmanned workshop, RRH1 and RRH2 are highly located and do the daily inspection. To improve the sensing accuracy, their received echoes are jointly processed in the BBU. In the manned workshop, workers are served by RRH4. The communication-only beam is used to meet their entertainment and interaction demands.  Regarding the unmanned vehicle, it is served by RRH3. The dual-function beams are used to schedule and monitor its actions. In this C-RAN, all of these signals are managed by the BBU, where an MEC is deployed for calculating the scheduling schemes and controlling the fabrication process. Moving a further step, the JCAS can be extended to joint sensing, communication, computing and control ($\mathbf{SC}^3$). They constitute $\mathbf{SC}^3$ closed loops 
that provide end-to-end solutions to different tasks. With effective feedback regime, the $\mathbf{SC}^3$ closed loop is like the reflex arc. Multiple loops can further form a strong machine nerve system that is promising to replace humans with all kinds of dirty, boring and dangerous works.

\subsection{JCAS with Dynamic 3D MIMO}
\label{subsec}
\begin{figure}[htbp]
	\centering	
	\includegraphics[width=0.5\textwidth]{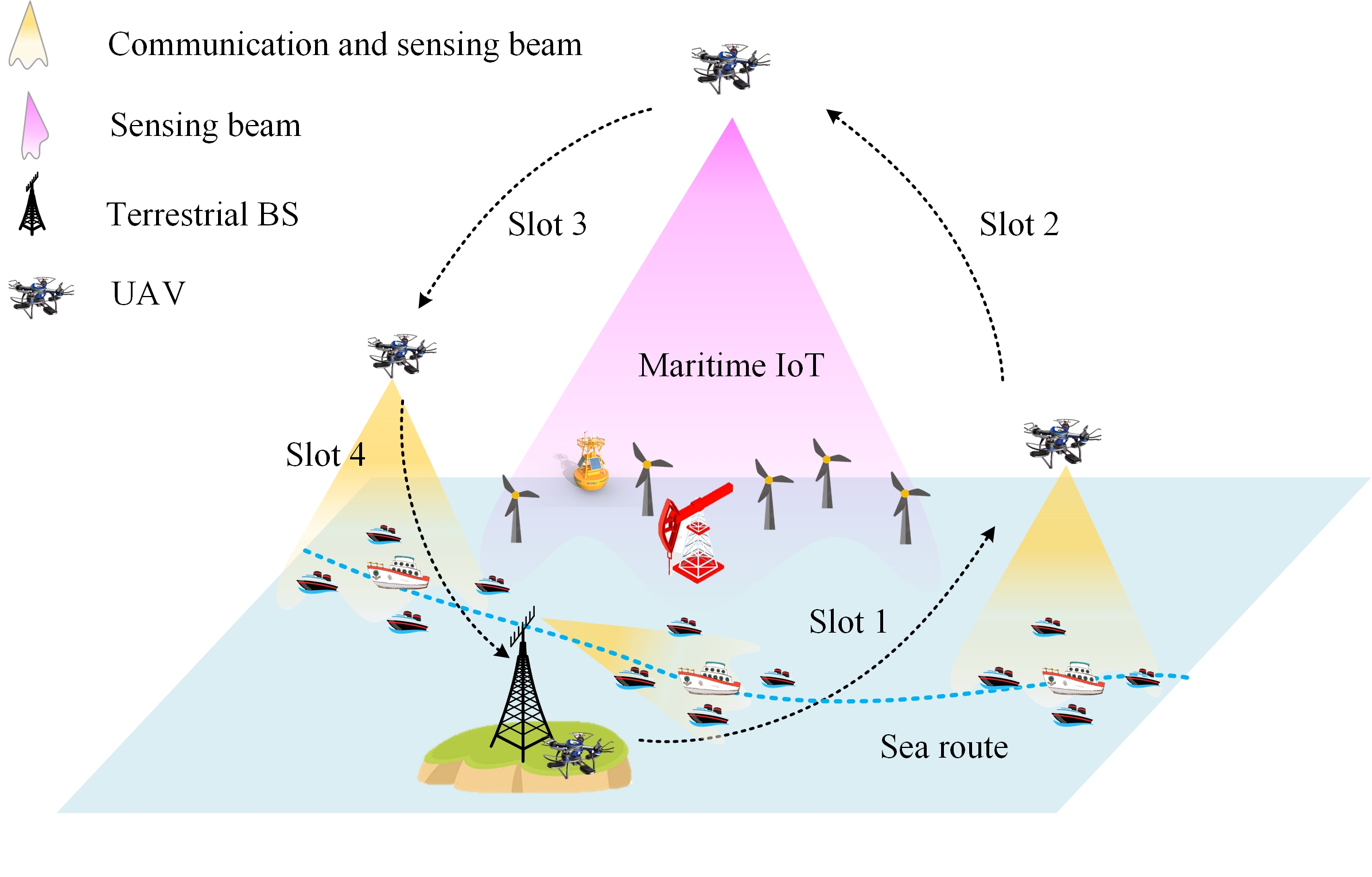}
	\caption{Illustration of a task executed by a UAV on the sea. The UAV uses different communication and sensing schemes in different time slots. The signaling strategy is adjusted with the trajectory of the UAV.}
	\label{fig3}
\end{figure} 

The fact that the antennas of terrestrial BSs are downward to cover ground users limits the BS view for sensing.  Aerial platforms, such as UAVs, airships, and even balloons, are necessary to provide complementary observations. In particular, by leveraging the maneuverability of UAVs, they could be flexibly deployed to provide on-demand services.  When the UAV flies high, the wide sensing beam could be used to illuminate the whole area, and when the UAV is close to the target or the user, the directional pencil beam could be used to refine the sensing resolution or improve the communication rate. Considering the whole flight of the UAV, the communication and sensing signals could be flexibly scheduled among different time slots. 
As shown in Fig. \ref{fig3}, we depict a task execution process of a UAV. In time slot 1, the UAV takes off according to a predefined route. It arrives at a given position and flies low to escort the fleet. The dual-function beam is used to satisfy the communication demand and simultaneously monitor the navigation environment. In time slot 2, terrestrial BSs are available to serve this fleet. The UAV thus lifts its altitude and uses wide sensing beams to patrol maritime IoTs. Then, in time slot 3, the UAV flies back and takes over the fleet from the terrestrial BS.
It serves this fleet until the fleet leaves out its jurisdiction. Then, the UAV returns to the ground, offloading the collected data and replenishing its energy. As we can see, this process requires the joint design of the UAV trajectory and the signaling strategy. But in reality, most UAVs cannot calculate the next-step strategy in a timely manner. Offline optimization is more practical by considering the energy limitations of UAVs.
In this sense, when we plan the actions of UAVs, timely CSI is not available \cite{107}. This makes the JCAS design under dynamic MIMO characterized by predictive and process-oriented traits \cite{103}. 

In the literature, Meng \emph{et al.} investigated a joint trajectory and radio scheduling scheme for the UAV-enabled communication and sensing integrated system, where the communication occupies the whole signal frame and the detection task only uses a proportion of the signal frame. Through joint optimizing the transmit precoding, the UAV trajectory and the sensing start time in each frame, the user rate was maximized under the constraint of the sensing beam pattern gain \cite{n6}. Lyu \emph{et al.} investigated a joint beamforming and static deployment or dynamic trajectory scheme. The problem is highly complex that the location/trajectory variables are exponent parts of the steering vectors. In return, the corresponding scheme owns great flexibility to balance communication and sensing by adjusting the beam pattern threshold \cite{f1}.
Moreover, UAV clusters support multi-scale sensing. This further improves the sensing resolution. In \cite{100}, the authors evaluated the performance of a cooperative sensing UAV network (CSUN), where UAVs simultaneously emit orthogonal beams for downward sensing and horizontal communication. A novel metric named the cooperative sensing coverage area was proposed and evaluated. Using this metric, the JCAS CSUN demonstrates a 66.3\% improvement compared with the communication and sensing separate CSUN.
In short, the JCAS with dynamic MIMO exploits DoFs in both the spatial and temporal domains. The communication and sensing functions are expected to be delicately arranged on the timeline and jointly optimized with the UAV's deployment. The joint optimization in both the spatial and temporal domains brings doubled DoFs but also makes the optimization complexity exponentially increased. The energy and hardware limitations of UAVs require the corresponding design to be simple. Compared to exhausting every DoF to chase the optimum, more robustness should be reserved to combat high dynamics.





\subsection{JCAS with Hybrid Active and Passive MIMO}

\begin{figure}[htbp]
	\centering	
	\includegraphics[width=0.45\textwidth]{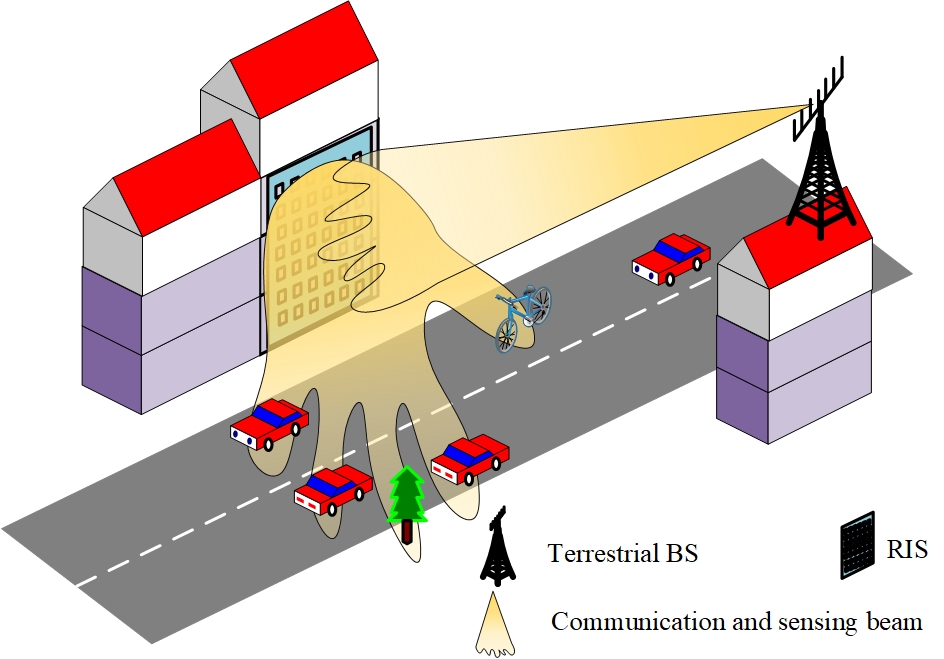}
	\caption{Illustration of an RIS-assisted communication and sensing waveform shaping. The RIS helps amplify the mainlobe of sub-beams and surpass the sidelobe between adjacent sub-beams. The primary webbed waveform is refined into a hand-like pattern by the RIS.}
	\label{fig4}
\end{figure} 
The JCAS system could embrace passive MIMO devices, namely RISs, to offer better communication and sensing services. Recent studies have shown that RISs have great potential in both the communication and sensing fields \cite{77}\cite{77s}. In the context of communication and sensing integration, the introduction of RISs provides more compatible solutions to integrate the communication and sensing functions.  As shown in Fig. \ref{fig4}, when the node only emits a low-quality waveform, the RIS helps to refine it. It adjusts its elements to amplify the mainlobes and suppress the sidelobes. As a result, the originally indistinguishable beam is reshaped into the one similar to a hand with distinguishable sub-beams. The target identifiability and user separability are largely improved.

In the literature, Wang \emph{et al.} applied an RIS to assist with DL communication. A joint active and passive beamforming scheme was proposed to minimize the inter-user interference. The waveform similarity  was  considered in the constraint \cite{78}. In \cite{116}, the authors further proposed a joint constant-modulus waveform and passive phase shift design, with the objective of inter-user interference minimization under the CRB constraint. Sankar \emph{et al.} adaptively divided the RIS into two parts: one group for communication and the other for localization. A multi-stage hierarchical codebook was designed to gradually refine location results while maintaining a good link to communication users \cite{79}.  
 Li \emph{et al.} maximized the weighted SNR of communication and sensing. The authors used manifold optimization to tackle the constant-modulus constraint of  RIS phase shifters. Since the probing signal experiences two-way transmissions between the target and the DFRC platform, RIS could bring four-fold gains of the received echoes while only tow folds for communication users \cite{r1}. 
Liu \emph{et al.} considered an integrated MIMO sensing and MU-MISO communication network and jointly optimized T\&R vectors and RIS phase shifters. 
An ADMM optimization framework based on the majorization-minimization (MM) scheme was developed to solve the formulated non-convex problem  \cite{r2}. 
Liu \emph{et al.} investigated the RIS in Terahertz bands. 
The authors used DRL combined with a primal-dual proximal policy to optimize the active and passive precoding. Simulation results verify the effectiveness of DRL to solve the high-dimensional non-convex problem \cite{r3}. 
In addition, Jiang \emph{et al.} invoked the RIS to enhance target echoes. The user rate requirement was considered in the constraint \cite{114}.  Yan \emph{et al.} and Song \emph{et al.}  maximized the radar SNR and the minimal illuminating power from the RIS toward different targets, respectively. The user SNR was applied as the communication metric and considered in the constraint \cite{r4}\cite{r6}. Wei \emph{et al.} applied multiple RISs to maximize the combined objective of the radar SINR and the minimal user SINR \cite{r7}. Hua \emph{et al.} minimized the transmit power subjected to the minimal communication and sensing SINR requirement and the cross-correlation requirement \cite{r8}.

In particular,  Sankar \emph{et al.} investigated the hybrid RIS that consists of both passive and active elements to communication and sensing integration. The transmit beamformer, RIS phase shifter and amplifier were jointly optimized to maximize the worst-case illuminating power toward different targets. Simulation results show that the hybrid RIS brings significant improvements compared with the passive RIS \cite{r5}.
Mishra \emph{et al.} considered the security issue of a multicast setting where the detected targets are potential eavesdroppers. The security rate was maximized subjected to the sensing SNR \cite{r10}.  Xu \emph{et al.} investigated an RIS-enabled backscatter communication and sensing system. In this system, the DFRC node sends signals for target tracking. These signals are also modulated by RISs for uplink data transmissions. These data is generated by  users who are executing computing tasks. The users can optionally upload raw data or computing results to the DFRC node. Under the above settings, the authors maximized a combined objective of sensing, communication and computing under practical time and energy constraints \cite{r11}.
He \emph{et al.} applied RISs to assist communication and sensing coexistence, and two RISs were used. The former is placed close to the communication transmitter to surpass the interference from the transmitter to the radar, and the latter is placed close to the communication receiver to surpass the interference from the radar to the communication receiver. The active and passive precoding matrices were jointly optimized to maximize the communication SNR with the constraint of the radar SNR \cite{n5}. 
More recently, Elbir \emph{et al.} did a comprehensive survey of RIS-assisted communication and sensing integration. This paper summarized that the RIS roles of coverage extension, interference suppression, and parameter estimation. It also pointed out that  environmental knowledge, waveform design, clutter and interference, and security are urgent open issues for RIS-assisted JCAS \cite{r9}. 

The above studies all show that the JCAS with active and passive MIMO delivers more satisfying communication and sensing performance compared with the no-RIS case.  However,  applying RISs to wireless communication is still open, let alone JCAS. It is not easy to make use of massive DoFs in a simple and robust way. To use these DoFs sensibly, 
the quantity of the prior knowledge shall match the quantity of DoFs. The acquisition of the data such as cascaded CSI, positions and the environment is new and challenging. A new architecture shall be established to collect and utilize these data. In addition, the joint optimization of active and passive beamforming is much more complex due to the constant modulus and discrete phase shift constraints of the RIS. 
Current approximated suboptimal solutions still have polynomial-level computing loads, which are far from what is actually bearable. One possible solution is to regard the RIS more as a part of the environment. They are not co-designed with active beamforming and only make changes in a coarse-grained manner.  For example, we could let the RIS alter its unit states based on a predefined codebook. Thus, the impinging beams would be radiated in different directions under different RIS patterns. Only when the output beam is directed to the target direction, is the received echo obviously amplified. Based on the amplitude differences of the received echoes, we can thus know the target direction.

\subsection{Discussion}
These new MIMO structures bring great opportunities to develop JCAS.  As for JCAS with cooperative MIMO, the flexible orchestration of multiple nodes and their functions is the key issue, where the complex cooperation and competition relationships among different nodes should be well addressed. In terms of the JCAS with dynamic 3D MIMO, the corresponding design tends to be process-oriented and predictive. Radio resource scheduling should be jointly considered with UAV's deployment. Joint active and passive MIMO brings great DoFs and also great complexity. Regarding RISs more as a part of the environment is friendly to actual deployment. In short, we find these JCAS MIMO models all enlarge optimization dimensions to earn more DoFs. To utilize these DoFs effectively, a new architecture such as a radio map is encouraged to be established \cite{104}. It is introduced to collect extrinsic data such as positions, the environment and the prior knowledge of targets. These data can be used to provide effective side information, e.g., large-scale CSI, for the JCAS optimization. Compared with the full CSI, the side information is predictable and slow-varying. 
The side-information-based schemes are thus executed on a new time scale. 
This is a middle time scale between the time-varying small-scale CSI and nearly static network deployment, so that provide a great balance between the cost and effectiveness.
By the way, the intelligent method is a good tool to bypass the theoretical obstacles and reduce the complexity of high-dimensional optimization. Effective neural network structures and objectives are expected to be found by trial and error in practice.

\section{JCAS in IoT scenarios}
 \begin{figure*}[tbp]
	\centering	
	\includegraphics[width=1\textwidth]{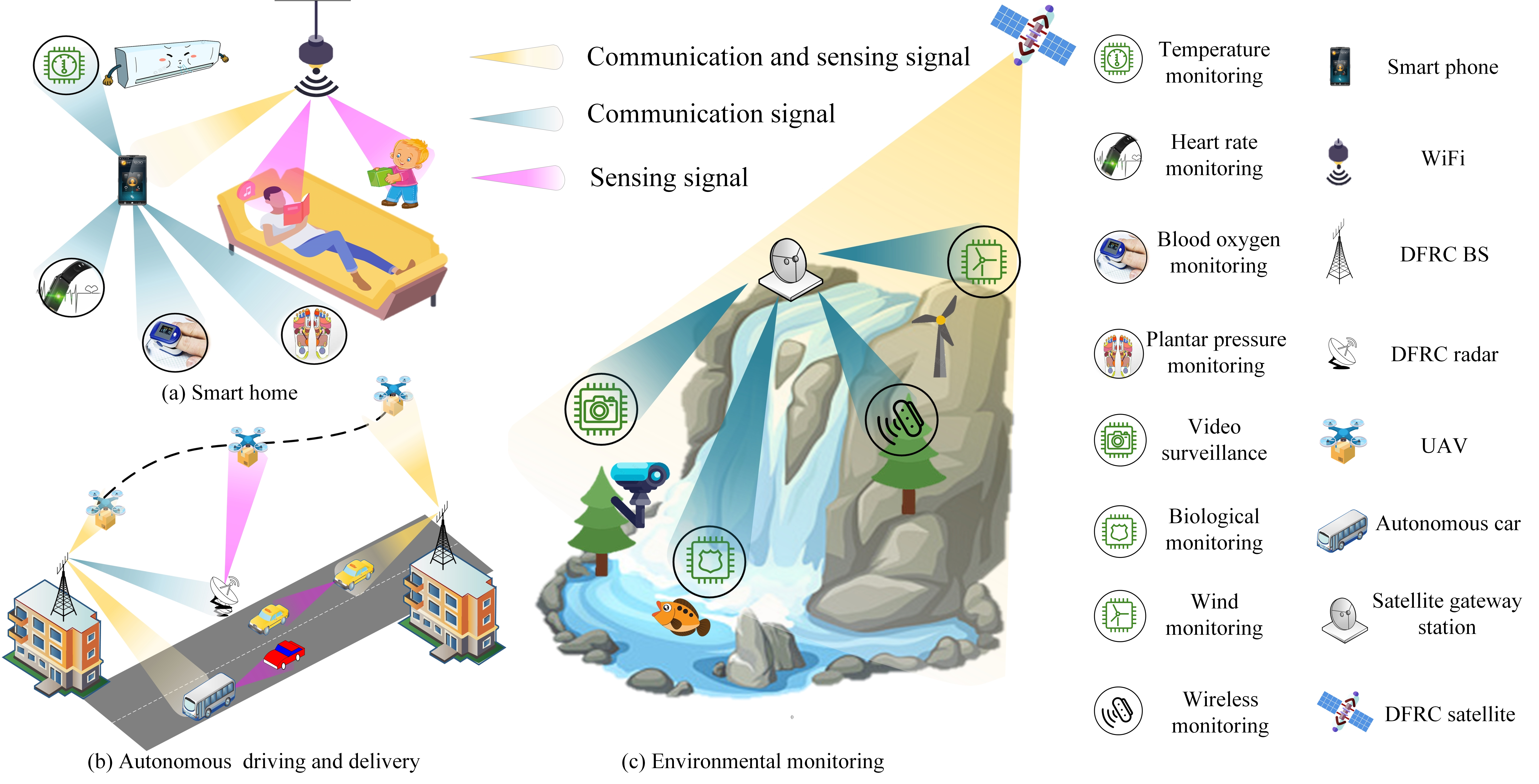}
		\caption{Illustration of the three typical applications of JCAS in IoT, i.e., smart home, autonomous driving and delivery and environmental monitoring.}
		\label{adfig4}
\end{figure*} 
 
\begin{table*}[t]
	
\begin{threeparttable}
	\caption{The JCAS schemes of ubiquity, green, complexity and cooperation issues}
	\label{table6}
	\begin{tabular}{|p{1.2cm}|p{3.4cm}|p{1.5cm}|p{1.5cm}|p{8.3cm}|}
		\hline
		
		\textbf{Issue}&\textbf{JCAS Type}\tnote{1}&\textbf{Setting}&\textbf{Waveform}&\textbf{Scheme}\\ 
		\cline{1-5}
		\multirow{14}{*}{ubiquity}&C\&S integration, novel waveform&ST-MIMO, SU-MISO&-&A hybrid beamforming scheme to maximize the energy efficiency with the waveform similarity as the constraint \cite{f2}.\\
		\cline{2-5}
		&C\&S integration, radar-centric&ST, SU-SISO&FMCW&Apply the CCS modulation when the elevation angle of the LEO satellite is low and switch to traditional modulations when the elevation angle exceeds a predefined threshold \cite{f3}. \\
		\cline{2-5}
		&C\&S integration, novel waveform&MT-MIMO, MU-MISO&-&A joint optimization scheme of the UAV trajectory, sensing start time and the transmit precoder to maximize the data rate with the guaranteed sensing beam pattern gain \cite{n6}.\\
		\cline{2-5}
		&C\&S integration, novel waveform&MT-MIMO, MU-MISO&-&A joint UAV deployment and beamforming scheme to maximize the weighted sum SINR of communication users with the guaranteed sensing beam pattern gain  \cite{f1}. \\ 
		\cline{2-5}
		&C\&S integration, novel waveform&MI-MIMO, MU-MIMO&OFDM&A novel antenna array design with an orthogonal C\&S beamforming scheme \cite{100}.\\  
		\hline
		\multirow{6}{*}{green}&C\&S integration, communication-centric&MT, SISO&OTFS&Preprocessing schemes to remove communication signals and parameter optimization to maximize the sensing SINR \cite{r17}.\\
		\cline{2-5}
		&C\&S integration, novel waveform&MIMO&-&Evaluation the beam synthesis scheme, the angular-of-arrival estimation scheme as well as  beam squint effects for the MBAA and a case study of a multi-beam JCAS framework \cite{f4}. \\
		\cline{2-5}
		&C\&S coexistence&MU-MISO&-&A cooperation scheme that jointly optimizes the time allocation of DL\&UL and active and passive beamforming \cite{r21}. \\
		\hline
		\multirow{8}{*}{complexity}
		&C\&S integration, radar-centric&SISO&PM-based LFM& A receiving sensing SINR maximization scheme that decouples the range and velocity estimation  \cite{f5}.\\
		\cline{2-5}
		&C\&S integration, communication-centric&ST-SISO&OFDM& A range and velocity decoupled algorithm \cite{f6}. \\
		\cline{2-5}
		&C\&S integration, communication-centric&ST, SISO&OFDM/OTFS&Preprocessing schemes of echo partition and adding virtual cyclic prefix and the optimal criteria of choosing the lengths of the subblock, the virtual cyclic prefix and the overlapping signal \cite{f7}.\\
		\cline{2-5}
		&C\&S integration, communication-centric&ST, SISO&OFDM&The overview of  pros and cons of OFDM sensing signals and an efficient decimation algorithm to reduce the redundancy of OFDM sensing signals \cite{f8}.\\
		\cline{2-5}
		&C\&S integration, communication-centric&MT, SISO&OTFS&A complex pattern-coupled sparse Bayesian learning scheme to estimate the velocity and range \cite{f9}. \\
		\cline{2-5}
		&C\&S integration, novel waveform&MT-MIMO, MU-MIMO&-&An iterative closed-form solution for joint radar signal covariance matrix and the communication physical precoder optimization \cite{63}.\\
		\hline
		cooperation&C\&S integration, networking&Multi-dual-function nodes, SISO&FMCW of sensing&A joint optimization scheme of C\&S power, communication time and quantization precision \cite{f14}. \\
		\hline
	\end{tabular}
\begin{tablenotes}
	\footnotesize
	\item[1] C\&S: communication and sensing.
\end{tablenotes}
	\label{Table6}
\end{threeparttable}

\end{table*} 

In recent years, we have witnessed the fast development of IoT. It was reported that in August 2022, the number of IoT terminals has exceeded the number of mobile phones in China. By interweaving sensors, actuators, and processors into a powerful ecosystem, IoT shows great potential to empower many novel applications. Since communication and sensing are two important pillars of IoT, it is natural to consider to apply JCAS to IoT. In Fig. \ref{adfig4}, we illustrate three typical IoT applications. They are smart home, autonomous driving and delivery, and environmental monitoring. It can be seen that thanks to JCAS, the isolation of communication and sensing is broken. The IoT systems have easy access to both communication and sensing resources and information. Such co-design framework enables both acute environmental sensing and convenient data exchange to empower these intelligent applications.

In this section, we consider the key requirements of JCAS in IoT. The discussions are from three different perspectives. The first is from the edge that BSs and radars provide communication and sensing services. Different from humans who are most in cities, IoT devices distribute much wider around the world. To provide ubiquitous services, it is necessary to extend ground techniques to airborne and spaceborne platforms. The second is from the end that IoT devices both have communication and sensing modules. The integration of communication and sensing would reduce their volume, cost, and energy consumption. But compared with edge infrastructures, their restricted hardware conditions limit the JCAS deployment.
Corresponding schemes are required to be green and simple. The third is from the network perspective. The cooperation of different nodes is the key to forming intelligent IoT systems that are capable of undertaking different complex tasks. 
For these reasons, we detail the issues of ubiquity, green, complexity, and cooperation for JCAS. In Table \ref{table6}, we summarize these schemes in terms of settings, waveforms, schemes, etc.

\emph{Ubiquity:} IoT devices are distributed around the world ubiquitously. To ensure that all the devices can obtain services anywhere, communication and sensing infrastructures shall be globally deployed. However, for remote areas, e.g., seas, mountains, and deserts, it is difficult to establish mesh terrestrial networks. Thus, the spaceborne and airborne JCAS schemes are expected to be investigated. Compared with the terrestrial scenario, the long transmission latency and large Doppler shift are two main challenges for the spaceborne JCAS. In the literature, Qiang \emph{et al.} investigated the integration of communication and sensing in the low earth orbit (LEO) satellite. Since the instant CSI is not available due to the long transmission latency, the authors used statistical CSI with additional attention to the beam squint effect instead. A hybrid beamforming scheme was developed to maximize the communication energy efficiency with the guaranteed radar waveform similarity \cite{f2}. Aliaga \emph{et al.} investigated the potential of differential absorption radar to use the chirp spread spectrum  (CSS) modulation for the data transmission. The radar is deployed in the LEO satellite and operates in the Terahertz band. Simulation results show that the CSS modulation has better performance than PSK when the evaluation angle of the satellite is low \cite{f3}.  As for the airborne platform, their flexible maneuverability provides additional DoFs to develop JCAS techniques\cite{n6,f1,100}. However, the large Doppler shift and limited onboard energy are two main restrictions in this case.

\emph{Green:} Apart from advanced IoT devices, there are many nodes that are low-cost, simply structured, and battery-powered. To prolong their lifetime, green techniques are of great importance to release the potential of JCAS in IoT. On the one hand, the JCAS technique itself shall develop a new branch that is green-oriented, i.e., energy-efficient dual-function waveform. Recently, Wu \emph{et al.} proposed an orthogonal time-frequency space (OTFS) waveform-based JCAS architecture. Compared with OFDM, the OTFS waveform is more energy-efficient due to its lower PAPR  \cite{r17}.
In \cite{f4}, the authors promoted the usage of the multi-beam antenna array (MBAA). Compared with massive MIMO, the MBAA uses the fixed analog beamformer which hardly consumes energy. Although fixing the analog part sacrifices some DoFs, the MBAA still owns great flexibility for the flexible beam synthesis. The corresponding design has the merit of low complexity thanks to removing the analog part. 
From another perspective, the combination of JCAS and green techniques, e.g., wireless power transfer \cite{r16} and backscatter communication \cite{r15} can improve the system sustainability. In addition, RISs could be widely deployed to provide energy-efficient MIMO solutions to assist IoT devices \cite{r18}. For example, Zhu \emph{et al.} recently proposed an RIS-assisted scheme for the coexistence of the sensor network and the cellular network. In particular, the sensors are equipped with a single antenna and use cooperative beamforming to transmit data. They are powered by the energy harvested from BS signals. 
This study shows that the RIS brings considerable gains to the network \cite{r21}. In short, the extended trade-off of communication, sensing, and energy is the key issue for the green JCAS.

\emph{Complexity:} To deploy JCAS ubiquitously in IoT nodes, it is important to lower its computing and hardware requirements.
As for the radar-centric scheme, Sahin \emph{et al.} developed a range and Doppler decoupled sensing algorithm for the PSK-based LFM radar waveform. The proposed algorithm overcomes the high complexity of joint range-Doppler processing \cite{f5}. 
As for the communication-centric scheme, Zeng \emph{et al.} proposed a range and speed separate estimation algorithm for the standard wireless communication waveform, e.g., IEEE 802.11ad and IEEE 802.11p. The proposed scheme has quite low complexity of only a one-dimensional fast Fourier transform \cite{f6}.  
In addition, Wu \emph{et al.} developed a low-complexity sensing algorithm using the OFDM waveform. The complexity is mainly from a Fourier transform \cite{f7}. Furthermore, the authors gave a comprehensive overview of the OFDM waveform. They revealed the redundancy of the OFDM sensing signals and proposed an efficient decimation algorithm to remove it. The proposed decimation-based sensing algorithm has lower complexity compared with the most efficient scheme to date \cite{f8}. 
In addition, Liu \emph{et al.} developed a parameter learning algorithm for OTFS-based radar waveform.
By exploiting the channel sparsity and the prior knowledge of motion limitations \cite{f9}, the complexity of the proposed  algorithm was largely reduced. 
However, the above schemes only tackled a single stream. 
It can not be directly extended to MIMO considering the antenna correlation. 
As for the novel waveform design, 
Dong \emph{et al.} investigated a beamforming scheme to jointly optimize the  radar signal covariance and the communication physical precoder. An iterative closed-form solution was derived to get rid of the optimization \cite{63}. 
To cope with the hardware complexity, Zhu \emph{et al.} devised a low-complexity MIMO-DFRC platform using low-resolution sampling. Based on the devised platform, the authors proposed a scheme to tackle the quantization distortion problem. It was shown to have 10\% of resolution grid accuracy with only 1-bit sampling \cite{f10}. In addition, Kumari \emph{et al.} established a fully-digital and dual-function platform, where the receiver only has few-bit analog-digital converters. This saves power  and also reduces the hardware volume \cite{f11}.

%
%

\emph{Cooperation:} Due to the cost and energy limitations, IoT devices are not standalone but cooperate with each other. 
A typical example is the sensor network, where a group of sensing nodes monitor the environment and send the data to a central unit for storage, analysis, and utilization. Similar to the cooperative MIMO, radio sensing devices could cooperate to send probing signals and process echoes. They form a large virtual array and exploit both micro and macro spatial DoFs. In addition, other types of sensors, processors, and actuators can also be included. Just like the nerve system, 
these nodes synergistically receive stimulus from the outside world, analyze the situation and make adaptations to the environment. Due to the high autonomy and intelligence, the $\mathbf{SC^3}$ framework has great potential to enable many cutting-edge applications as in Fig. \ref{adfig4}. In \cite{f12}, Feng \emph{et al.} extended JCAS to joint communication, sensing and computing (JCSC). The authors showed how these three functions can assist with each other and proposed a unified JCSC machine-type framework. As for autonomous driving, Liu \emph{et al.} proposed to abstract the resources of communication, computing, and control and jointly manage them in a fine-grained manner. The authors showed that such co-design can better support safety-critical applications than the separate optimization \cite{f13}. In addition,
Wen \emph{et al.} considered an edge inference task. In the considered setting, distributed dual-function nodes use time division multiple access to perform the radio sensing, feature calculation, and feature uploading, successively. A task-oriented framework was established to jointly optimize the sensing power, communication power and time, and the quantization resolution. It was shown that the co-design framework brings considerable gains compared with the separate counterpart \cite{f14}. 

\emph{Discussion:} Applying JCAS to IoT is to shape generalized schemes into tailor-made solutions for specific settings. Given the traits of IoT, the ubiquity, green and complexity issues are expected to be addressed.  By considering the dynamic environment and restricted hardware conditions, the schemes that sacrifice efficiency for low complexity, low energy consumption, and high reliability are encouraged to be developed. But how to achieve the high efficiency-cost ratio while guarantee the communication and sensing performance is still open. A theoretical breakthrough is expected to quantitatively analyze the JCAS schemes in terms of cost and efficiency. 
Moving a further step, the power of IoT lies in the cooperation of different nodes. The JCAS takes the first step to deepen the cooperation from the data level to the signal level. In the future, it is expected to break the isolation of not only communication and sensing modules but also computing and control modules. The $\mathbf{SC^3}$ co-design provides an end-to-end solution for the supported tasks. 
To enable $\mathbf{SC^3}$ integration, the key is to derive the task-oriented objective that theoretically depicts the relationships of $\mathbf{SC^3}$. This objective may be different for different kinds of tasks.  Obviously, this needs a long-term effort.

\section{Open issues and Future Directions}
In this section, we briefly outline open issues and promising directions for JCAS. As the research on communication and sensing integration has just started, there is still great uncertainty on its future development. However, one can expect further works on intelligence and security. One could also envision the interplay between JCAS and other cutting-edge technologies to take advantage of their mutual benefits.
\subsection{JCAS in Integrated Space-Air-Terrestrial Network}
To extend the coverage of both communication and sensing, it is necessary to design JCAS in the space-air-ground integrated network (SAGIN). In this scenario, the distinct rate, latency and reliability of satellite, aerial, and terrestrial links would render new challenges. Two kinds of integration, i.e., communication and sensing integration and space-air-ground integration, could couple with each other. This consequently poses great challenges to the system design. One possible solution is to explore the hierarchical architecture of the hybrid system. As shown in \cite{126}, one may derive basic models for satellite-terrestrial cooperation and treat a complicated hybrid system as the combination of basic models. On this basis, the basic JCAS-SAGIN model is a great breakthrough point to analyze  complex JCAS-SACINs. Each basic model contains both minimal space-air-ground infrastructures and minimal communication and sensing functions. Thus, the basic relationships of different platforms and functionalities are kept in these models. The agile orchestration of these basic models would lead to various large-scale JCAS-SAGINs. In this direction, both theoretical analysis and key technologies require research attention.
\subsection{JCAS  Using Artificial Intelligence}
%
Applying artificial intelligence (AI) to JCAS may reduce the computing burdens of the MIMO design. 
A trained network could directly output the JCAS schemes by giving the input raw data. In addition, although the learning process is still a black box without explicit explanations, the output policies may be heuristic for the theoretical analysis. However, it is not an easy task to extract high-level information from massive raw data. The effective reward/cost objectives and neural network structures remain to be studied.
Perhaps the  model-based methods and data-based methods shall be combined. For example, if we  know the output results are sparse based on the prior knowledge of the physical system, the neural network structure could be sensibly designed to simplify the training process.

\subsection{Joint Sensing, Communication, Computing and Control}

Future 6G networks are envisaged to shift from connecting things to connecting intelligence. In other words, we want to endow connected machines with human-like intelligence. To do so, the network needs to be aware of device status and instruct their behavior. In this sense,
JCAS that makes wireless networks perceptive is the first step. We shall further investigate to build a ``nerve  system" for machines, where a closed loop of $\mathbf{SC}^3$ is a basic unit similar to a reflex arc.
Under such closed loops, numb machines can adapt to the environment and accomplish different tasks automatically, thus releasing humans from all kinds of dangerous and boring jobs.
However, optimization over $\mathbf{SC}^3$ closed loops covers information theory, estimation theory, control theory, etc.  A unified theoretical model is required to figure out the basic relationship of $\mathbf{SC^3}$.

\subsection{Security Issues of JCAS}
Security is one of the key issues for JCAS. The sensing function requires signals to fully interact with the environment so that the surrounding information could be imprinted in the waveform. This increases the risk of being eavesdropped.
In addition, unlike communication users, who are authenticated before access, targets are not identified and are more likely to be malicious.  How to illuminate the target while limiting the information leakage is still open. Furthermore, the JCAS would bring explosive data. 
The balance of data security and data efficiency is challenging. We may combine JCAS with the blockchain technique to establish a distributed open network. The explosive data are stored in different nodes and exchanged, updated and cleared through the blockchain. In this way, the whole life of the data is recorded. But the information synchronization cost of the whole network is huge. The balance of efficiency and security is an interesting topic for future study.  

\subsection{Combination of JCAS, Backscatter Communication and RIS}
JCAS could combine other techniques, such as ambient backscatter communication and RISs, to further exploit the new usage of the electromagnetic wave. Different from JCAS, which focuses on the signal processing in T\&R nodes, ambient backscatter communication and  RISs change the signal in the process of propagation. Thanks to the properties of low cost, RISs and backscatter tags can be ubiquitously deployed. In this way, they are able to make the random environment under control. An intelligent environment may be further constructed to better support the JCAS. 
However, the basic relationship of communication, sensing and the environment is unknown. Research on the design of JCAS with a smart radio environment is still in its infancy.

\section{Conclusions}


In this paper, we have reviewed recent advances in JCAS MIMO. Detailed schemes of communication and sensing coexistence and communication and sensing integration have been presented. We have also investigated  three novel JCAS MIMO models combined with cutting-edge technologies. In these JCAS schemes, we have found that MIMO mainly plays the role of directional beamforming for the communication function and waveform shaping for the sensing function. The main challenges lie in using restricted DoFs to balance their incompatible interests. Targeted at the dimensional problem of using MIMO, we have discussed possible solutions based on simple and robust principles. Afterwards, we have specified JCAS in IoT scenarios and emphasized the issues of ubiquity, green, complexity and cooperation. On this basis, open issues have been outlined, with a great vision to embrace a ubiquitous, intelligent and secure JCAS network in the upcoming 6G era.

\end{document}